\documentclass[aps,prx,twocolumn,superscriptaddress,citeautoscript,floatfix,showpacs,longbibliography]{revtex4-1}

\usepackage[english]{babel}
\usepackage[utf8]{inputenc}
\usepackage{amsmath}
\usepackage{graphicx}
\usepackage[dvipsnames]{xcolor}
\usepackage{amssymb}
\usepackage{anysize}
\usepackage{amsmath}
\usepackage{physics}
\usepackage{makeidx}
\usepackage{float}
\usepackage{xcolor,colortbl}
\usepackage{listings}
\usepackage{enumitem}
\usepackage[group-separator={,}]{siunitx}
\usepackage{subfigure}
\usepackage{caption}
\usepackage{lipsum}
\usepackage{qcircuit}
\usepackage{xr}
\usepackage[draft]{pdfcomment}

\usepackage[paperwidth=210mm,paperheight=297mm,centering,hmargin=2cm,vmargin=2.5cm]{geometry}
\usepackage{natbib}

\usepackage[all]{hypcap}
\providecommand{\newoperator}[2]{\newcommand*{#1}{\mathop{\mathrm{#2}}\nolimits}}
\newoperator{\sgn}{sgn}
\newoperator{\arctanh}{arctanh}
\newoperator{\argmax}{argmax}
\newoperator{\diag}{diag}

\def\bra#1{\langle#1|}
\def\ket#1{|#1\rangle}
\def\braket#1#2{\langle#1|#2\rangle}

\begin{document}
\title{Unitary Block Optimization for Variational Quantum Algorithms}

\author{Lucas Slattery}
\author{Benjamin Villalonga}
\author{Bryan K. Clark}

\affiliation{Department of Physics and IQUIST and Institute for Condensed Matter Theory, University of Illinois at Urbana-Champaign, IL 61801, USA}
\begin{abstract}
Variational quantum algorithms are a promising hybrid framework for solving chemistry and physics problems with broad applicability to optimization as well.
They are particularly well suited for noisy intermediate scale quantum (NISQ) computers.
In this paper, we describe the unitary block optimization scheme (UBOS) and apply it to two variational quantum algorithms: the variational quantum eigensolver (VQE) and variational time evolution.
The goal of VQE is to optimize a classically intractable parameterized quantum wave function to target a physical state of a Hamiltonian or solve an optimization problem.
UBOS is an alternative to other VQE optimization schemes with a number of advantages including fast convergence, less sensitivity to barren plateaus, the ability to tunnel through some local minima and no hyperparameters to tune. 
We additionally describe how UBOS applies to real and imaginary time-evolution (TUBOS).

\end{abstract}
\maketitle

\section{\label{sec:intro}Introduction}
In the near term, quantum computers remain limited in qubits and state coherence. These early noisy intermediate-scale quantum (NISQ) computers cannot perform error correction, making many promising quantum algorithms unusable \cite{preskill_quantum_2018}.  To avoid these issues,  hybrid classical-quantum algorithms like the Quantum Approximate Optimization Algorithm (QAOA) and the Variational Quantum Eigensolver (VQE) leverage the resources of a quantum computer to simulate and sample from a classically intractable state while using classical resources to limit the qubit and coherence requirements of a problem \cite{farhi_quantum_2014,peruzzo_variational_2014}.

VQE takes as input a problem Hamiltonian, $\hat{H}$, and a variational family of parameterized quantum circuits, $\ket{\Psi(\boldsymbol{\theta})}$, and attempts to find the set of parameters  $\boldsymbol{\theta}=\arg\!\min_{\boldsymbol{\theta}} E(\boldsymbol{\theta})$ which minimizes the energy $E(\boldsymbol{\theta})=\bra{\Psi(\boldsymbol{\theta})}\hat{H}\ket{\Psi(\boldsymbol{\theta})}$.  VQE accomplishes this by iteratively computing operator expectation values (e.g. energy gradients) of $|\Psi(\boldsymbol{\theta})\rangle$ on the quantum computer and then classically updating the control parameters $\boldsymbol{\theta}$
until a convergence criterion is met.  The VQE algorithm must run many copies of a circuit with limited depth.
The standard approach to VQE has been improved in various ways including better initialization, error mitigation techniques, and measurement minimizing protocols 
\cite{cerezo_cost-function-dependent_2020,grimsley_adaptive_2019,lee_generalized_2019,barkoutsos_improving_2020,grant_initialization_2019,omalley_scalable_2016,kokail_self-verifying_2019,zhu_training_2019,liu_differentiable_2018,huggins_towards_2019,hempel_quantum_2018,gard_efficient_2020,lee_generalized_2019,zhou_quantum_2020,shen_quantum_2017,benedetti_parameterized_2019,li_efficient_2017,sagastizabal_experimental_2019,kandala_error_2019,mcardle_error-mitigated_2019,verteletskyi_measurement_2020,izmaylov_unitary_2020,gokhale_on3_2020,huggins_efficient_2021,izmaylov_unitary_2020,rubin_application_2018,kandala_hardware-efficient_2017,jena_pauli_2019,yen_measuring_2020,gokhale_minimizing_2019,rubin_application_2018}.   

In this work, our focus will be on improving the approach to circuit optimization.  Early VQE optimization used costly gradient free classical optimization methods including particle-swarm optimization (PSO) and the Nelder-Mead (NM) method \cite{peruzzo_variational_2014,shen_quantum_2017,wecker_progress_2015,zhu_training_2019,benedetti_parameterized_2019}. Since then, most VQE optimization proposals rely either on some form of gradient descent in order to optimize the variational parameters \cite{grimsley_adaptive_2019,kubler_adaptive_2020,mitarai_quantum_2018,kandala_hardware-efficient_2017,nannicini_performance_2019,grimsley_adaptive_2019,wang_accelerated_2019,parrish_quantum_2019,foss-feig_holographic_2020,hempel_quantum_2018,nakanishi_subspace-search_2019,liu_differentiable_2018,huggins_non-orthogonal_2020,sagastizabal_experimental_2019,ryabinkin_constrained_2019} or a sampling of quadrature points for finding parameter updates \cite{nakanishi_sequential_2020,parrish_jacobi_2019}.
Gradients can be computed either from finite differences or analytically using quantum circuits. 
Although formally a hybrid quantum-classical algorithm, VQE with gradient descent involves minimal classical computation beyond averaging observables and adding the gradient onto the current set of parameters.  In the context of VQE, gradient descent has several known challenges including local minima, significant hyperparameter tuning, slow convergence, and 
exponentially vanishing gradients (i.e. the barren plateau problem) \cite{mcclean_barren_2018,cerezo_cost-function-dependent_2020,zhou_quantum_2020, zhu_training_2019}. 

In this paper, we overcome many of these challenges by providing and benchmarking a flexible, hyper-parameter free optimization algorithm for VQE optimization, the unitary block optimization scheme (UBOS).
We also describe TUBOS, the application of UBOS to the problem of time evolution.
UBOS sweeps over gates optimally minimizing the energy of one gate at a time in the environment of the other temporarily fixed gates.  Our approach has a number of advantages including requiring significantly fewer operator expectation values to converge, an ability to tunnel through some local minima and a robustness to barren plateaus which comes from avoiding the direct use of the exponentially small gradients over the circuit parameters. In addition, it off-loads a non-trivial amount of computational work to classical machines more equally balancing the classical aspect of quantum-classical hybrids.     

We demonstrate UBOS using as a variational ansatz,
\begin{equation}
\left |\psi \right \rangle = \prod_{j=1}^{K} U_j \ket{0}
\label{ansatz_eq}
\end{equation}
obtained by applying K generic two qubit unitaries $U_j \in SU(4)$ (i.e. quantum gates) acting on adjacent qubits.  We will assume these gates are laid out in a brick pattern (see Figure \ref{fig:ansatz}) with a gate-depth $d$ although non-local two-qubit gates can be straightforwardly used as well.

The rest of the paper is organized as follows. In Section~\ref{sec:methods}, we describe how to implement the UBOS for VQE.
Next, in Section~\ref{sec:implementation}, we demonstrate the method on the 1D XXZ Heisenberg model comparing against stochastic gradient descent.
We explore the method's performance both with exact expectation values from the circuits as well as values which are correct on average but stochastically noisy as would result from a finite number of measurements. 
We also analyze UBOS' ability to avoid certain barren plateaus.
Finally, in Section~\ref{sec:VITE} we describe TUBOS and demonstrate its efficacy for variational imaginary time evolution. 
We conclude the paper in Section~\ref{sec:discussion}  with a discussion of our main results.

\section{\label{sec:methods}Unitary Block Optimization}

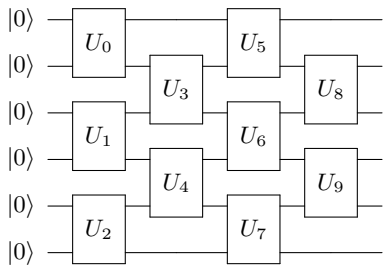
\begin{figure}[b]
\begin{tabular}{c}
        \Qcircuit @C=1em @R=1em {
        \lstick{\ket{0}} & \multigate{1}{U_0} & \qw & \multigate{1}{U_5} & \qw & \qw \\
        \lstick{\ket{0}} & \ghost{U_0} & \multigate{1}{U_3} & \ghost{U_5} & \multigate{1}{U_8} & \qw \\
        \lstick{\ket{0}} & \multigate{1}{U_1} & \ghost{U_3} & \multigate{1}{U_6} & \ghost{U_8} & \qw \\
        \lstick{\ket{0}} & \ghost{U_1} & \multigate{1}{U_4} & \ghost{U_5} & \multigate{1}{U_{9}} & \qw \\
        \lstick{\ket{0}} & \multigate{1}{U_2} & \ghost{U_4} & \multigate{1}{U_7} & \ghost{U_{9}} & \qw \\
        \lstick{\ket{0}} & \ghost{U_2} & \qw & \ghost{U_7} & \qw & \qw
        }
\end{tabular}
\caption{
Structure of our generic unitary circuit ansatz for 6 qubits and 4 layers.
}
\label{fig:ansatz}
\end{figure}

In this section, we describe UBOS using generic two-qubit unitaries. The generalization to any k-qubit unitary is straightforward.  One can also apply UBOS to common hardware gates, including the two-parameter fSim gate, (see Appendix \ref{sec:AppA}) and to restricted manifolds of larger unitaries (see Appendix \ref{sec:large_unitaries}) with significantly fewer quantum measurements.
A generic (local or non-local) two qubit unitary, $U_j$, can be represented as a linear combination of 16 two qubit Pauli strings,
\begin{equation}
U_j = \sum_{\alpha,\beta=0}^{3} t_j^{\alpha\beta} P^{\alpha\beta}
\label{paulidecomp}
\end{equation}
where $P^{\alpha\beta} = \sigma^{\alpha} \sigma^{\beta}$ and the $t_j^{\alpha \beta}$ are constrained to preserve the unitarity of $U_j$. Consider a depth-d unitary circuit with $K$ unitary two qubit gates, $\ket{\psi} = \Pi_{j=0}^{K}U_{j}\ket{0}$. The derivative of $|\psi\rangle$ with respect to $t_j^{\alpha\beta}$, $\ket{\psi_{j}^{\alpha\beta}} \equiv |\partial \psi/\partial t_j^{\alpha\beta}\rangle $,  is given by
\begin{equation}
 \ket{\psi_{j}^{\alpha\beta}}= \left(\prod_{k=1}^{j-1} U_k \right) P^{\alpha\beta} \left( \prod_{k=j+1}^{K} U_k \right) \ket{0}
\end{equation}
which is the result of substituting gate $U_j$ by the Pauli operator $P^{\alpha \beta}$, and so is also a circuit of depth $d$.

The energy $E(\bf{t})$ is a function of the parameters $\bf{t} \equiv \{\bf{t_1},\bf{t_2}...\bf{t_K} \}$ where  $\mathbf{t_j} = (t_j^{00},t_j^{01},\dots\,t_j^{33})$.  If we fix the parameters for all but the $j$'th gate, the energy can be written as 
\begin{equation}
\begin{aligned}
E(\bf{t_j}) &  =  \sum_{\alpha,\beta,\alpha,\beta'}^{3} t_j^{*\alpha'\beta'}t_j^{\alpha\beta} \Tilde{H}^{\alpha\beta;\alpha'\beta'}\\
 & = \mathbf{t_j}^{\dagger} \Tilde{H} \mathbf{t_j}
\label{EtHt}
\end{aligned}
\end{equation}
with $\Tilde{H}$ having the matrix elements
\begin{equation}
\label{Htilde}
\begin{aligned}
    \Tilde{H}^{\alpha\beta;\alpha'\beta'} = \bra{\psi^{\alpha\beta}_j}   H \ket{\psi^{\alpha'\beta'}_j}
     \end{aligned}
\end{equation}

By expanding $H$ into the sum of unitary operators, we can compute each of the 136 unique matrix elements of $\tilde{H}$ (16 real and 120 complex) by a set of circuits of depth at most $2d$ (see Fig. \ref{fig:gradmeasurecircuit}).  
An alternative approach for computing $\tilde{H}$ using circuits of only depth $d$ via solving for the unknown $\tilde{H}_{ij}$ in Eq.~\ref{EtHt} is given in Appendix \ref{sec:HfromEs}.
Given $\Tilde{H}$ the computation of the lowest energy state for gate $j$ (with all other gates fixed) can now be computed classically by minimizing Eq. \ref{EtHt} with respect to $\bf{t_j}$ under the unitary constraints. This is a 16 parameter optimization problem that can be solved using any classical technique including gradient descent, parallel tempering, Nelder-Mead, etc.  This approach can take a much larger step in parameter space than gradient descent as well as tunnel through local minima to find the global minimum for this gate. 
In fact, without considering stochastic noise, an UBOS step (per gate) down in energy is always at least as large as an SGD step.

After finding the optimal $\bf{t_j}$ for $U_j$, UBOS updates its parameters and proceeds similarly with all other unitaries in a predetermined order; we have chosen to randomly shuffle the updating order every sweep. An interesting open question is to determine if there are better sweeping orders.  A full sweep of UBOS happens after each gate has been updated once. 

UBOS can be run in a highly parallel fashion with every circuit needed to evaluate $\tilde{H}$ able to be simultaneously computed. 
In parallel, one can evaluate each matrix element of $\tilde{H}$, each Hamiltonian Pauli-term for each matrix element, and each repetition to accumulate statistics for each Pauli-term.

Each circuit of UBOS can be further accelerated compared to the naive circuit construction using various acceleration approaches which have been introduced for optimizing VQE circuits by SGD.
One can analytically cancel out some gates with their inverse further in the circuit. For example, for a nearest neighbor brick-pattern ansatz (i.e. see Fig. \ref{fig:ansatz}), the total number of gates used by any one UBOS circuit scales as $O(d^2)$ and is independent of the system size $L$ when $L \gg d$.  Note that this cancelling allows UBOS to optimize any system size $L$ VQE circuit with a quantum computer which only has $2d$ qubits even if $L\gg d$.   Finally, UBOS can be improved by using the many proposed methods for reducing the amount of measurements needed by grouping operators into commuting groups  \cite{verteletskyi_measurement_2020,fizmaylov_revising_2019,gokhale_on3_2020}. 

The cost (in expectation values) of running an entire sweep of UBOS is comparable to performing 8.5 (all-gate) gradient descent steps.  This follows as gradient descent requires computing (per gate $j$) the 16 expectation values $\bra{\psi} \hat{H} \ket{\psi_{j}^{\alpha\beta}}$ via a depth $2d$ circuit (see Appendix \ref{sec:AppB}). UBOS will therefore be more efficient in cases where it reaches the same energy as SGD with, at least, an order of magnitude less sweeps. UBOS is also similar to other proposed gradient-free gate optimizers that optimize subsets of parameters \cite{nakanishi_subspace-search_2019,parrish_jacobi_2019}.  These methods instead require a number of measurements which scale exponentially with the number of independent parameters and whose classical optimization is both analytical (as opposed to the classical optimization of UBOS) and require exponential (with parameters) classical resources.  For example, optimizing a $U(4)$ gate requires $O(16^2)$ expectation values in UBOS compared to $3^{15}$ expectation values (and classical resources) in these other approaches. 

UBOS bears resemblance to a number of classical variational approaches including DMRG \cite{white_density_1992} and the linear-method in variational Monte Carlo \cite{umrigar_optimized_1988}.  In both cases, one evaluates an  effective Hamiltonian $\Tilde{H}$ in the tangent space of derivatives.  This (sometimes generalized) eigenvalue problem on the effective Hamiltonian is then solved and used to find a new set of parameters.
In UBOS, if we relax the unitary constraints on $\bf{t_j}$, minimizing the energy becomes an eigenvalue problem $\Tilde{H} {\bf{t_j} }= E \Tilde{S} \bf{t_j}$ where $\Tilde{S}^{\alpha,\beta;\alpha'\beta'}=\braket{\psi^{\alpha\beta}}{\psi^{\alpha'\beta'}}$.  $\Tilde{S}$ is not necessary for UBOS but can be computed if desired with 16 additional expectation values. In UBOS, the solution to the eigenvalue problem gives a lower bound on the constrained optimization problem and these unconstrained parameters could be used to initialize the classical optimization.  The unitary constraint is necessary because, while the solution to the unconstrained problem gives a valid wave-function, it cannot be represented by a single unitary circuit.

\section{\label{sec:implementation}Application to 1D XXZ Heisenberg Model}

We benchmark UBOS using a classical simulation of a quantum computer on the 1D XXZ Heisenberg Hamiltonian:
\begin{equation}
    \hat{H}=\sum_{j}\sigma_{j}^{z}+\sum_{j}\sigma_{j}^{z}\sigma_{j+1}^{z}+\sum_{j}(\sigma_{j}^{x}\sigma_{j+1}^{x}+\sigma_{j}^{y}\sigma_{j+1}^{y})
    \label{xxz_1d}
\end{equation}
and compare it with an analytical stochastic gradient descent method.  See Appendix \ref{sec:AppB} for details on expectation value measurements for both methods.
\subsection{Noiseless Simulations}
\begin{figure}
\includegraphics[width=\linewidth]{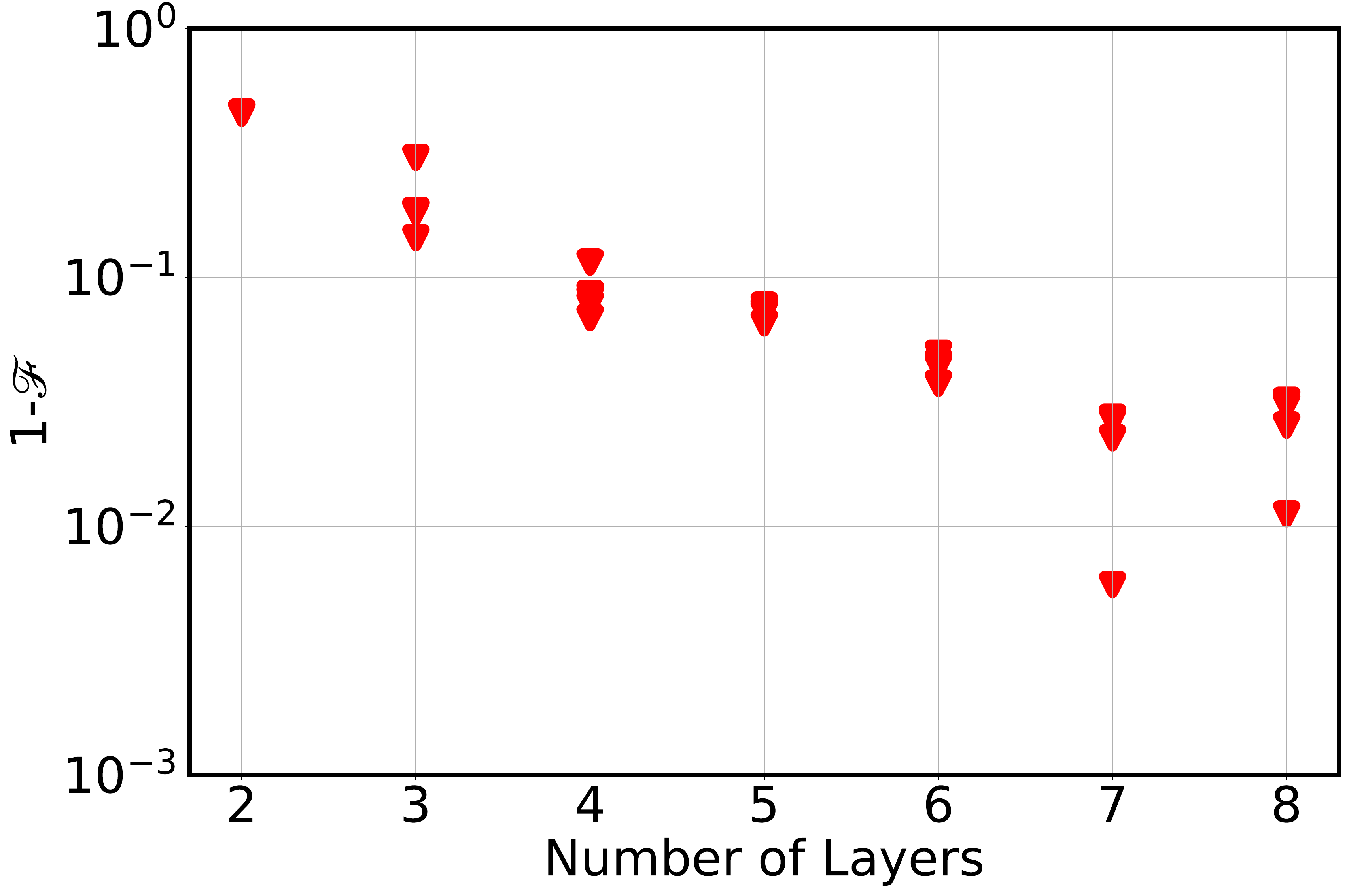}
\caption{Infidelity of ground state vs ansatz depth for a 12 site XXZ 1-D Heisenberg model (Eq. \ref{xxz_1d}). Each point represents a UBOS run from a randomly initialized ansatz. For each of the 12 site ansatz we ran until plateaued.}
\label{fig:FvsLayers}
\end{figure}

\begin{figure*}[htb]%
\onecolumngrid
\centering
\subfigure[][]{%
\label{fig:14n7layers-a}%
\includegraphics[width=0.48\linewidth]{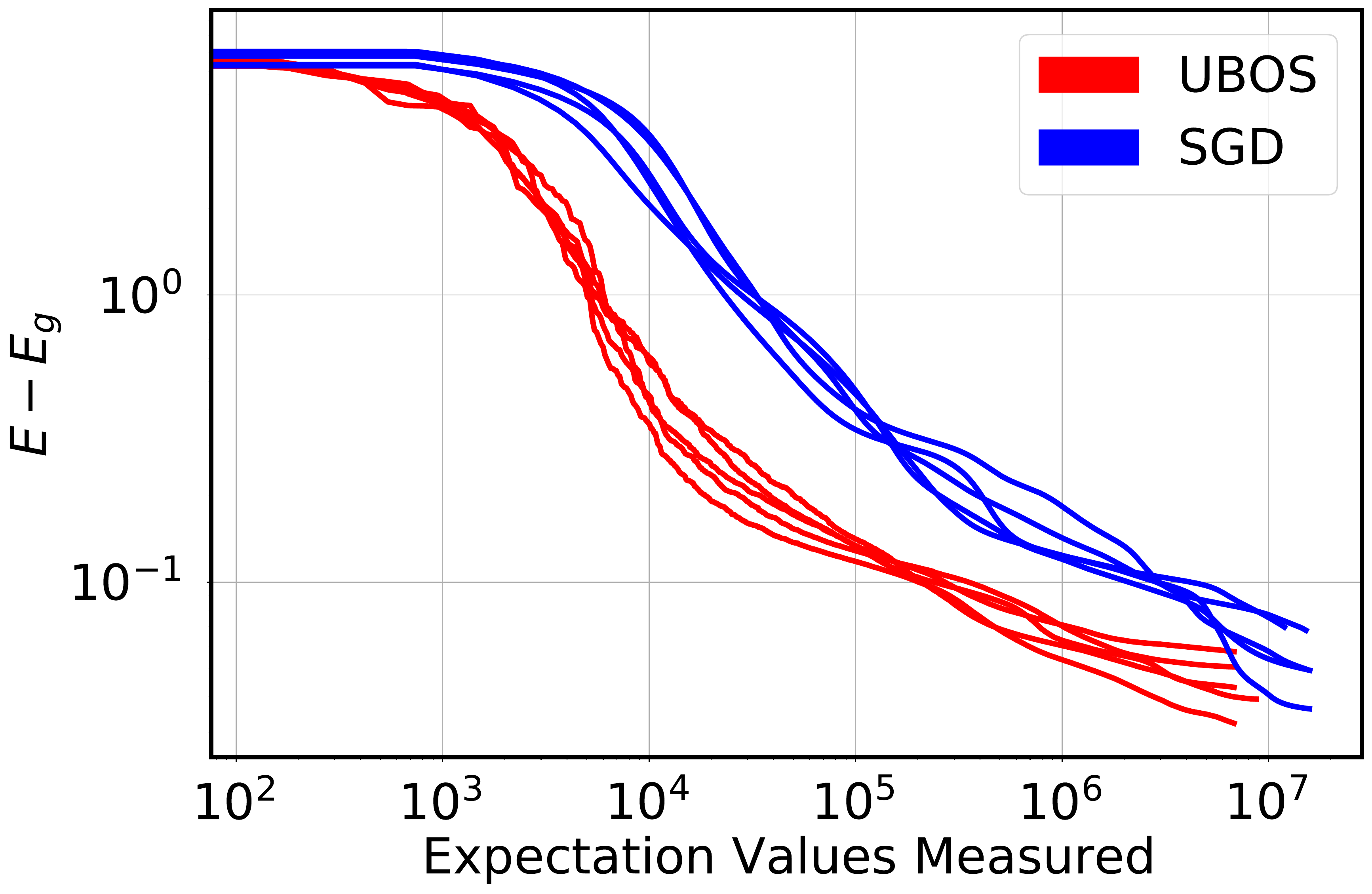}}%
\hspace{8pt}%
\subfigure[][]{%
\label{fig:14n7layers-b}%
\includegraphics[width=0.48\linewidth]{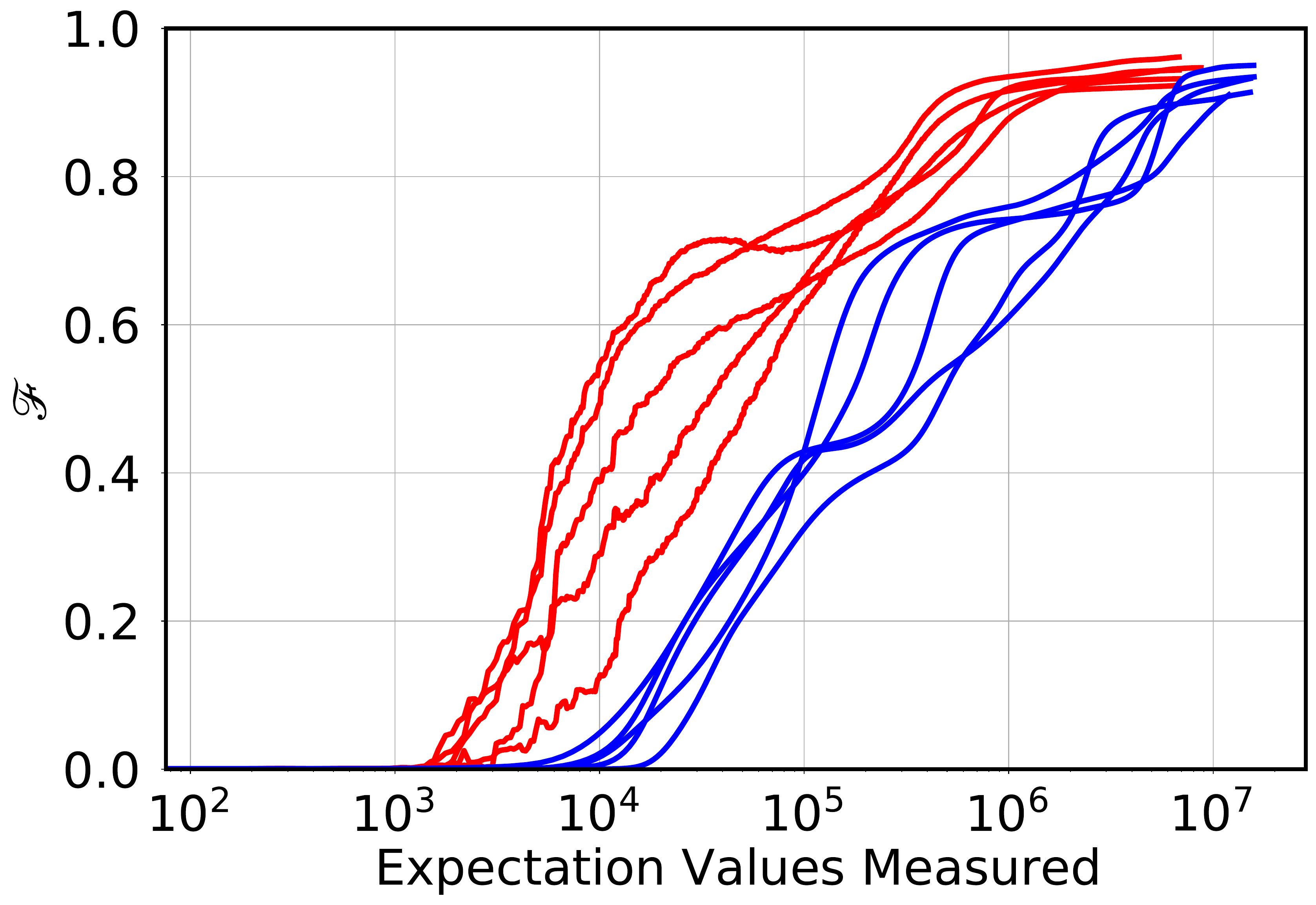}}%
\caption[]{Prototypical noiseless runs for a 14-qubit 7-layer ansatz for UBOS and SGD  showing  \subref{fig:14n7layers-a} Ground state energy difference and  \subref{fig:14n7layers-b} Ground state fidelity vs number of expectation values. 
Each UBOS line is a different initial starting circuit but the SGD and UBOS are both (pairwise) initialized to the same starting circuits.
}%
\label{fig:14n7layers}%
\end{figure*}

\begin{figure}
    \centering
    \includegraphics[width=\linewidth]{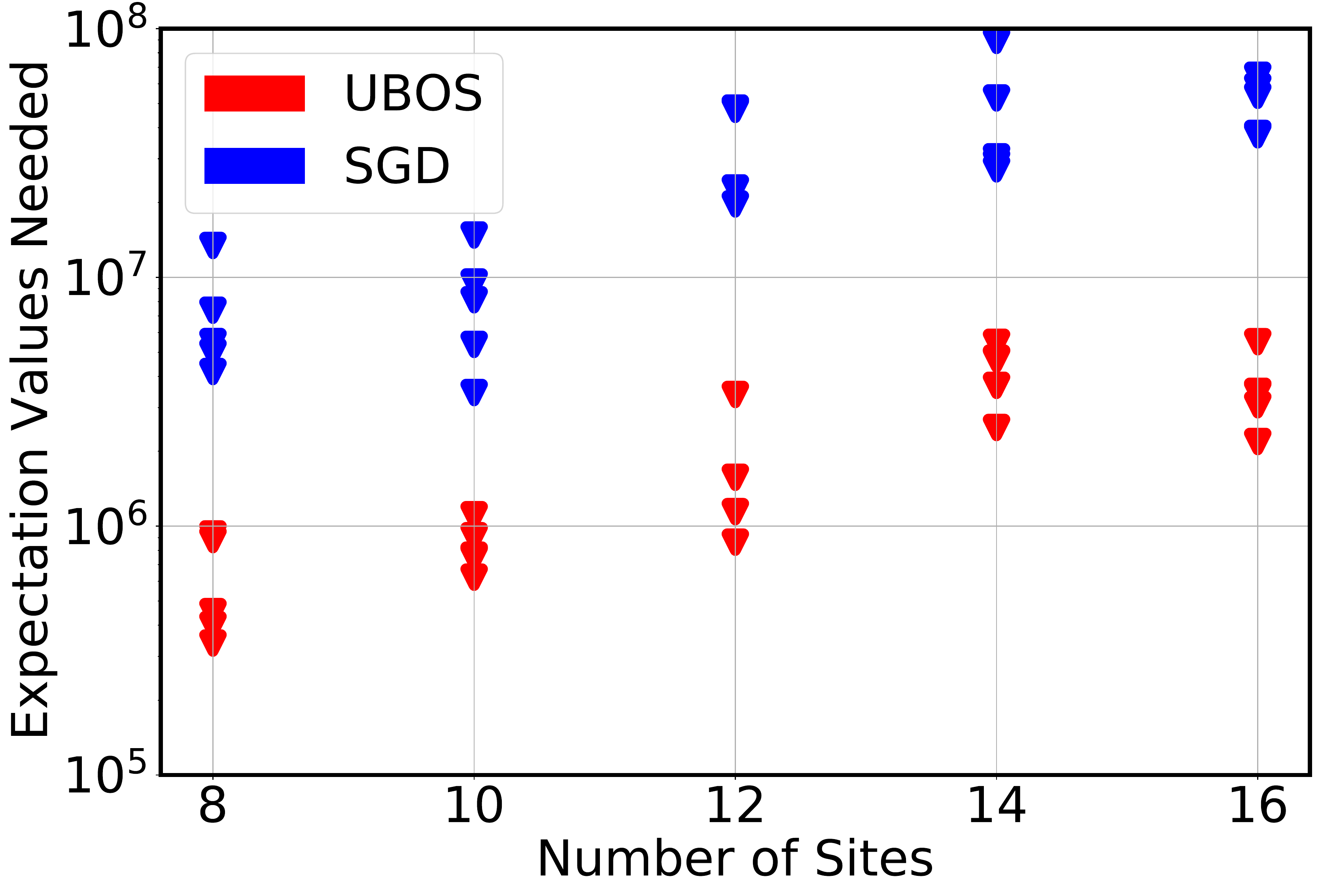}
    \caption{Number of expectation values needed to reach an energy error per site of $10^{-2}$ for a 7 layer ansatze with varying number of sites. Each point represents a different single optimization run (with pairwise matching starting conditions between UBOS and SGD).}
\label{fig:evn_vs_sites}
\end{figure}

We first benchmark UBOS assuming the expectation values computed by the quantum computer are exact and noiseless.  We emulate this classically by contracting a tensor network version of the relevant circuits using the TensorNetwork library \cite{roberts_tensornetwork_2019}.   
The ansatz's two qubit unitary blocks are parameterized with the KAK decomposition \cite{tucci_introduction_2005}.  Each gate in our initial circuit is generated randomly by selecting the KAK parameters uniformly at random from $[0,\pi)$.  
While given enough classical resources, one can always find the global minimum of a 16 parameter optimization,  in practice we devote enough resources in our classical optimization to find values close but not always equal to this true global minimum at each step.
For the comparison with SGD, we compute the gradient of all gates and take a step downhill of $5\%$ of the gradient, which we experimentally found to work well. 
We performed numerical experiments at different system sizes and ansatz depths.  Under UBOS optimization, increasing layers to the VQE ansatz gives exponential improvement in the infidelity $1-\mathcal{F}$, where $\mathcal{F} = |\langle \Psi_\textrm{exact}| \Psi_\textrm{VQE}\rangle|^2$  and $|\Psi_\textrm{exact}\rangle$ and  $|\Psi_\textrm{VQE}\rangle$ are the exact ground state and the VQE approximation respectively (see Figure \ref{fig:FvsLayers}).
For all scenarios studied, both optimization methods reached similar final fidelities, although we were unable to fully plateau many of the optimization runs. In Figure \ref{fig:14n7layers}, we show representative optimization runs for a 14 site, 7 layer ansatz.  Throughout the optimization, UBOS reaches a given energy or fidelity with approximately an order of magnitude less expectation values measured (see Appendix \ref{sec:AppC} for details).  We proceed to look at the expectation values needed to reach an energy difference per site of $10^{-2}$ for a seven layer ansatz as we change the number of sites from 8 to 16 (see Figure \ref{fig:evn_vs_sites}).  Here we also consistently find UBOS requires an order of magnitude fewer expectation values.   We also considered doing both UBOS and SGD on one gate at a time, finding it was significantly more efficient to do UBOS (see Appendix~\ref{sec:AppD}). 

\subsection{Simulations with Gaussian Noise}

\begin{figure*}[htb]%
\centering
\subfigure[][]{%
\label{fig:runswvariance-a}%
\includegraphics[height=142pt]{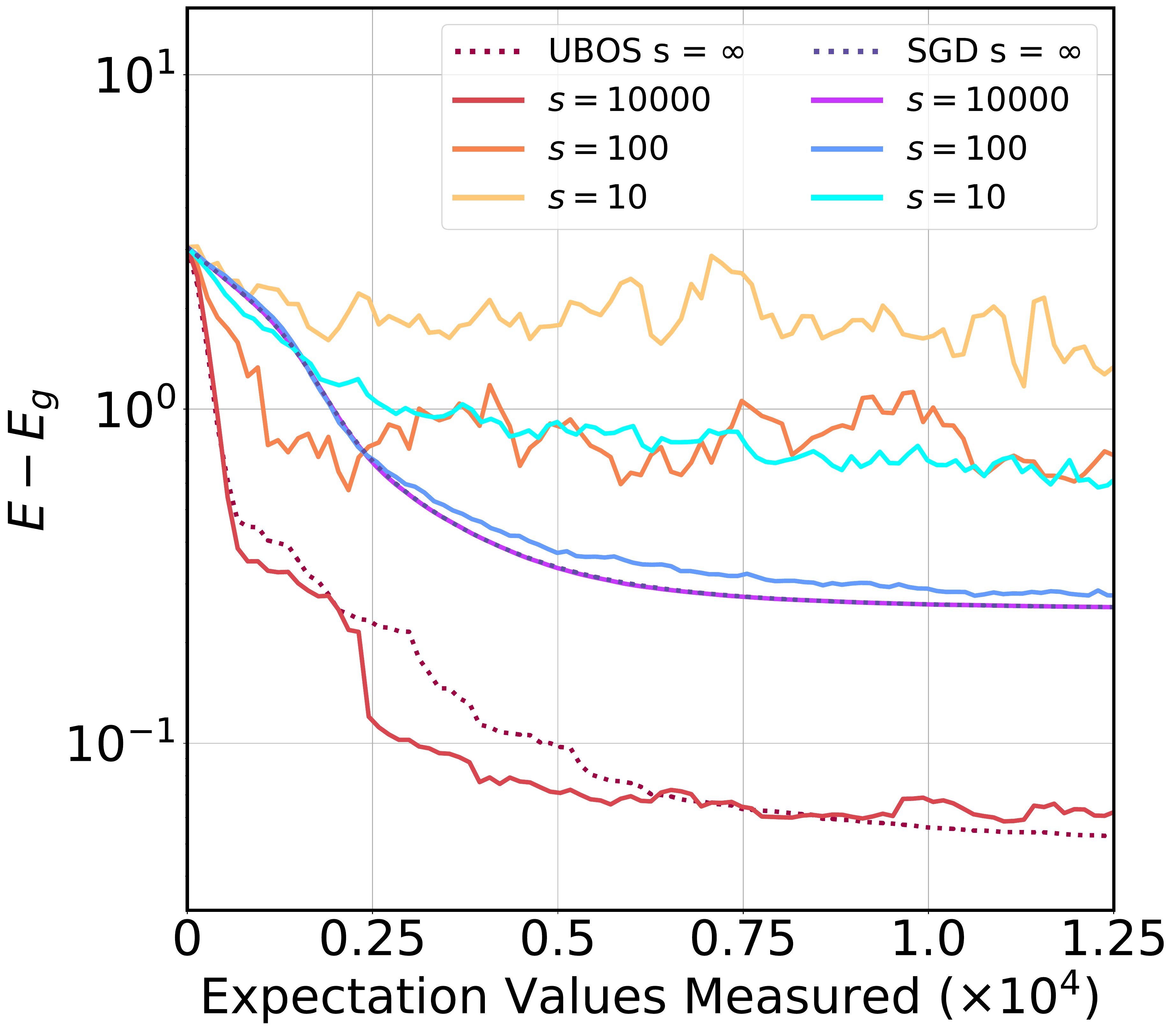}}%
\hspace{2pt}%
\subfigure[][]{%
\label{fig:runswvariance-b}%
\includegraphics[height=142pt]{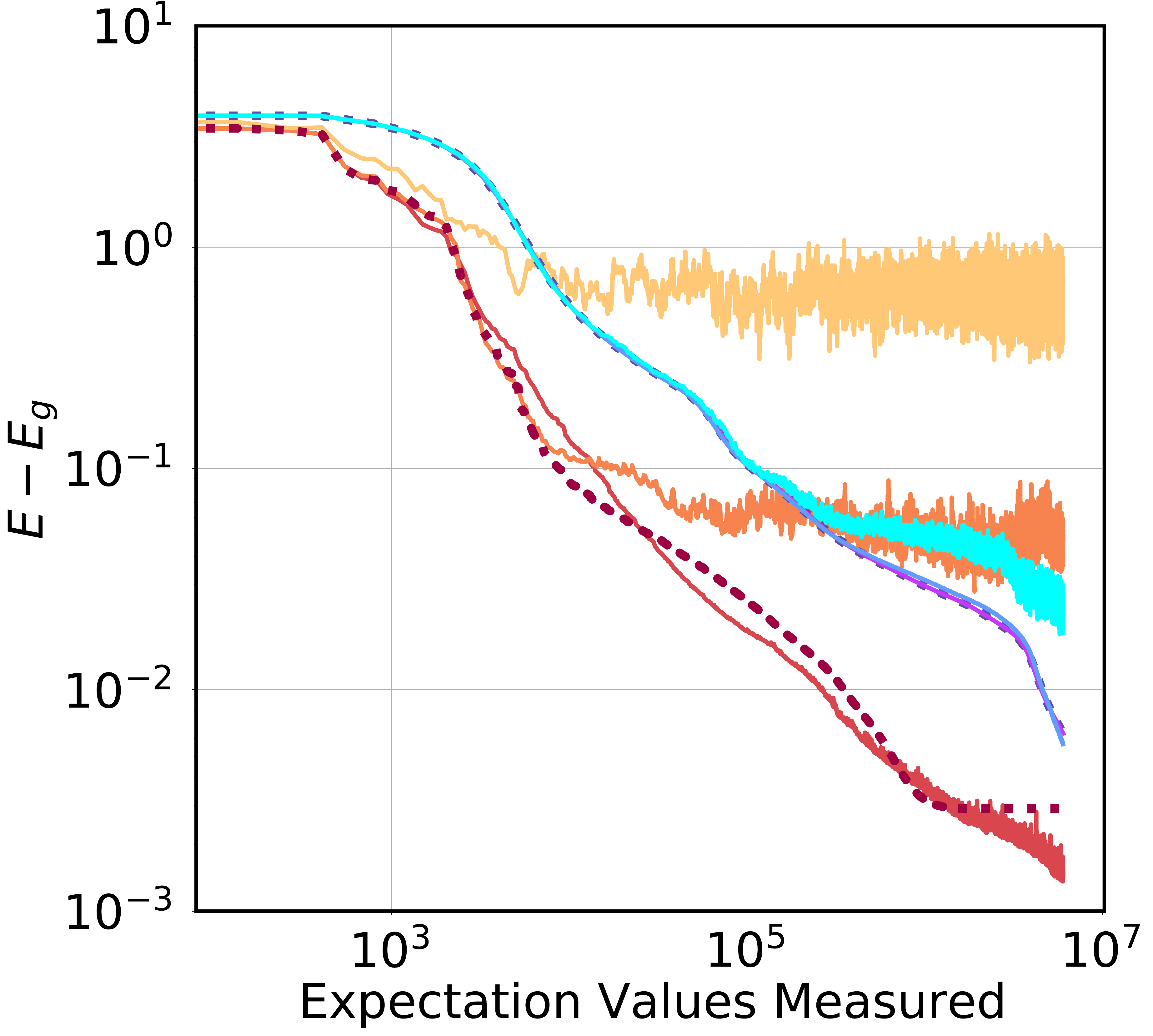}}%
\hspace{2pt}%
\subfigure[][]{%
\label{fig:runswvariance-c}%
\includegraphics[height=142pt]{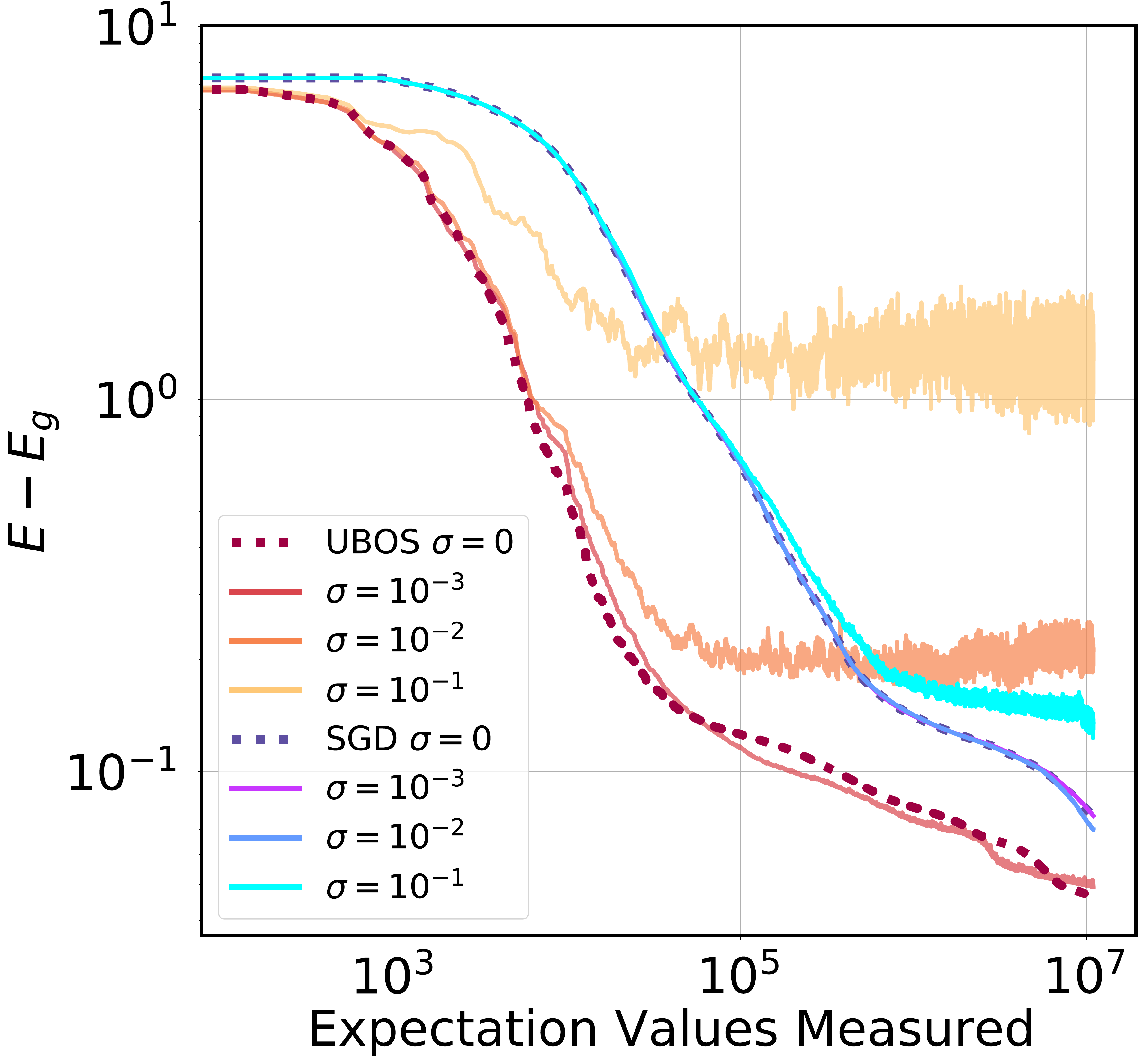}}%
\caption[]{Comparing the behavior of the optimization methods with statistical noise.  Energy difference from the ground state vs number of expectation values computed.
\subref{fig:runswvariance-a} Simulations using IBM's Qiskit (on classical emulators) for a 6 site, 3 layer ansatz with varying number of samples $s$ per operator measurment.
\subref{fig:runswvariance-b}  Simulations for different gaussian noise $\sigma$ for 8 site, 7 layer ansatz optimizations and \subref{fig:runswvariance-c} 16 site, 7 layer ansatz optimizations.  When $\sigma = 10^{-3}$, the noisy SGD optimization  is not visible as it overlaps with the corresponding exact optimizations.}%
\label{fig:runswvariance}%
\end{figure*}

In order to better understand the performance of UBOS with noisy expectation values, as would happen with limited samples of the circuit, we perform two tests.  In our first test, we directly evaluate the effect of noise coming from a finite number of measurements by implementing the relevant circuits for UBOS in Qiskit and performing  simulations (without gate noise) on a 6 site 3 layer ansatz for the 1D XXZ Heisenberg model \cite{Qiskit}.
We find that very noisy estimates of $\tilde{H}$ cause the UBOS algorithm to plateau at higher energies than the exact, noiseless UBOS simulation (see Fig.~\ref{fig:runswvariance}(a)). SGD seems less sensitive to this stochastic noise, where we find even taking 100 samples per gradient achieves a similar plateau as the exact case (although the case of 10 samples per gradient is unclear).   Interestingly, though, the exact case gets stuck in a local minima plateauing significantly above the exact UBOS simulation.

\begin{figure*}[htb]%
\centering
\subfigure[][]{%
\label{fig:onestep-a}%
\includegraphics[width=0.33\linewidth]{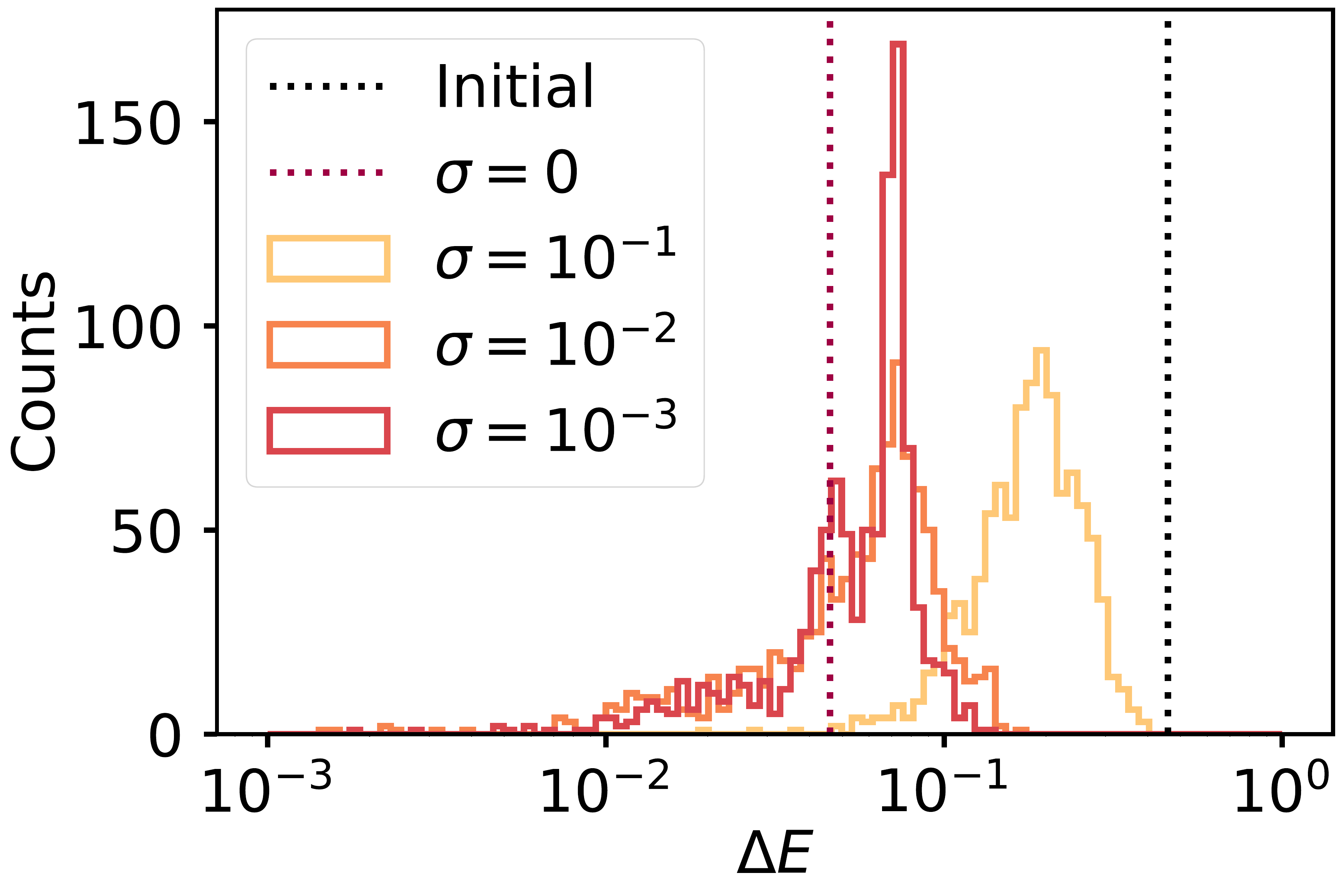}}%
\hspace{2pt}%
\subfigure[][]{%
\label{fig:onestep-b}%
\includegraphics[width=0.33\linewidth]{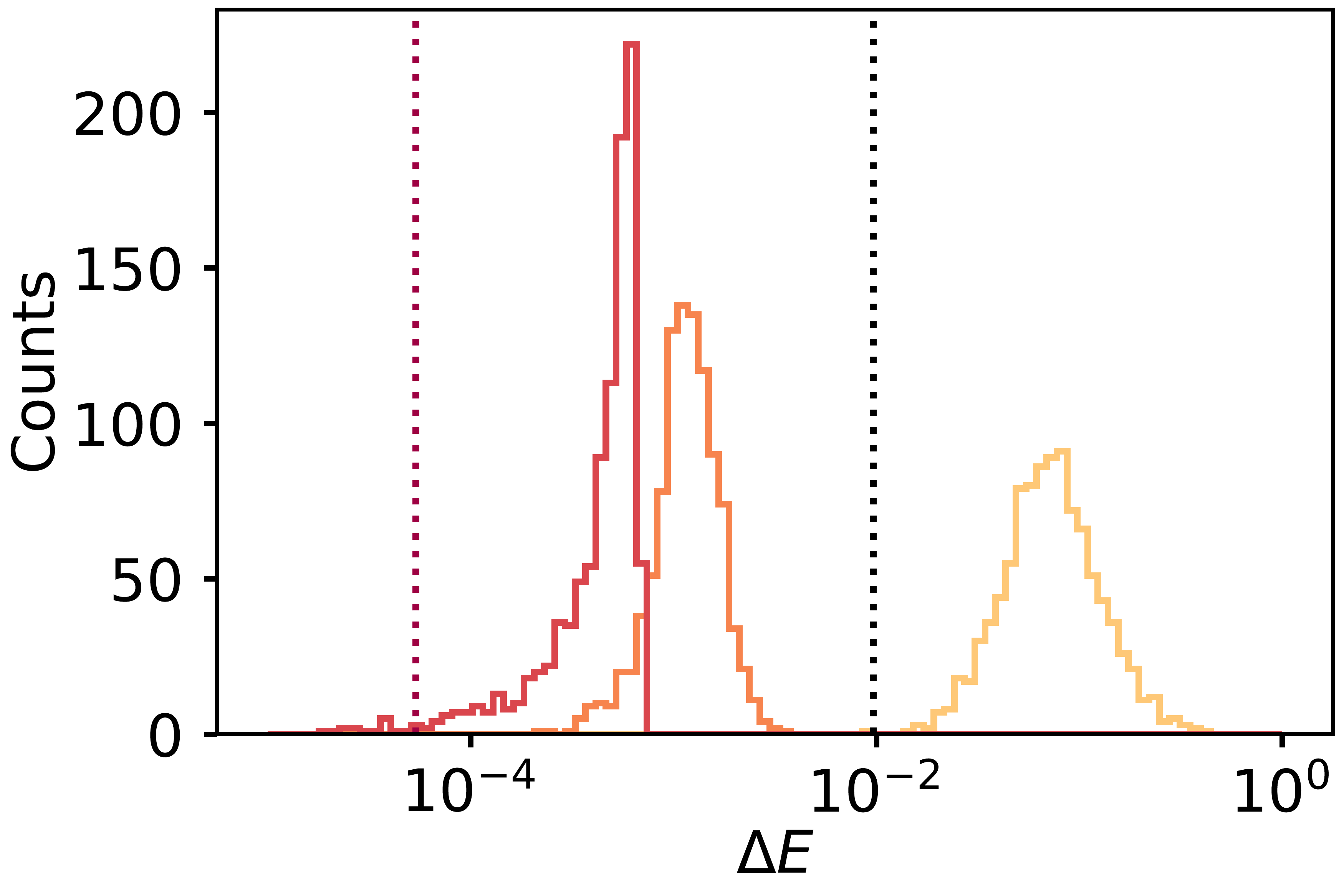}}%
\hspace{2pt}%
\subfigure[][]{%
\label{fig:onestep-c}%
\includegraphics[width=0.33\linewidth]{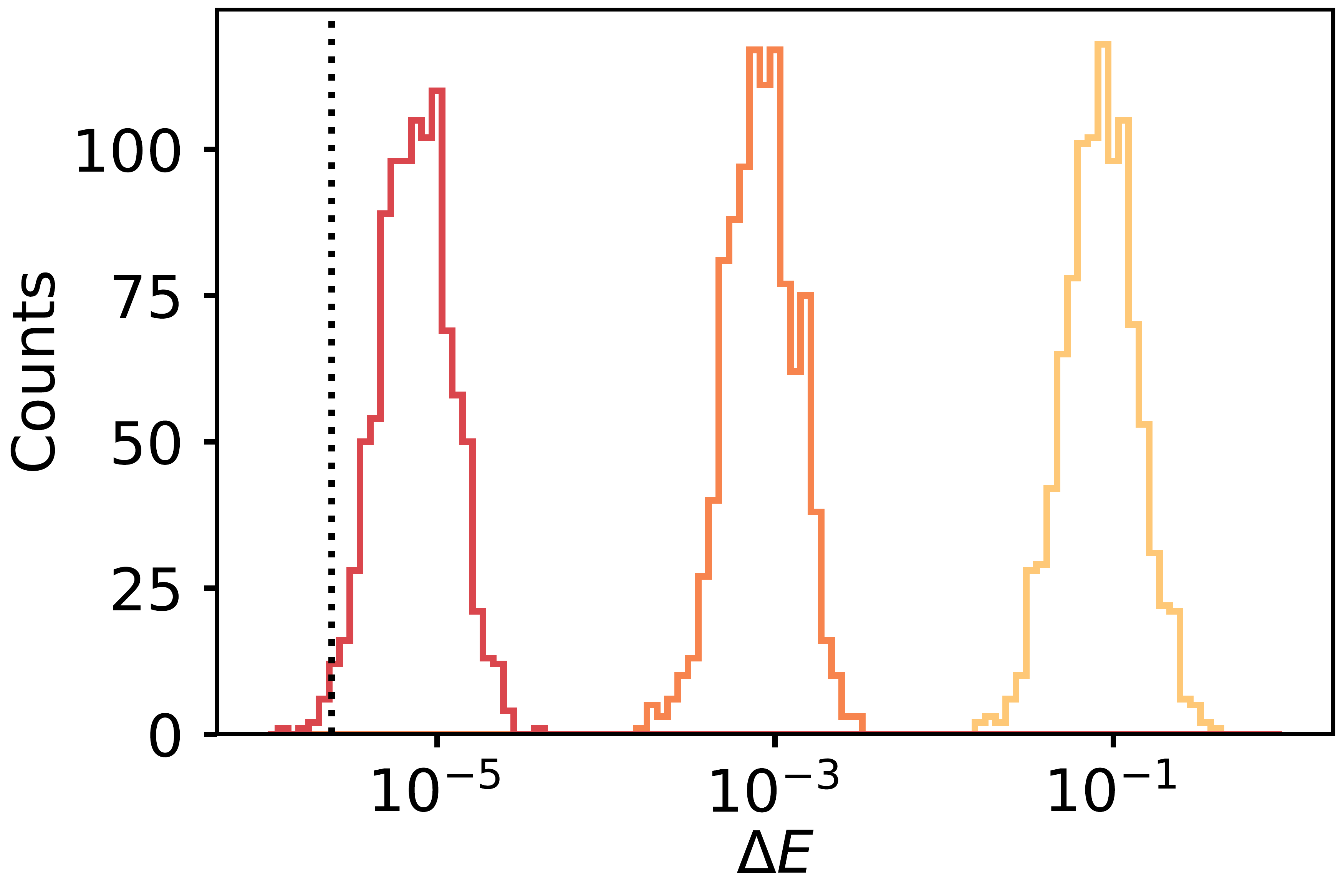}}%
\caption[]{Energy difference $\Delta E$ from the lowest energy seen during optimization of a single UBOS step performed on a gate in the middle of a 16 site, 7 layer ansatz showing the energy before optimization (black dotted line), the energy after noiseless optimization (red dotted line) and the distribution of energies for optimization with added Gaussian noise with standard deviation $\sigma$ per matrix element, for 1000 realizations. The noiseless optimization doesn't always have $\Delta E=0$ because of imperfect classical optimization using Nelder-Mead.  UBOS step shown  \subref{fig:onestep-a} at the beginning of optimization 
\subref{fig:onestep-b} after 100 optimizations. 
\subref{fig:onestep-c} after 1,000 optimizations. 
}%
\label{fig:onestep}%
\end{figure*}

We further validated this behavior on larger systems (8 and 16 site, 7 layers) by computing the exact expectation values and explicitly adding Gaussian noise with width $\sigma$ to each matrix element of $\tilde{H}$ (while keeping it Hermitian) finding similarly that UBOS saturates at a noise dependent energy (see Figs.~\ref{fig:runswvariance}\subref{fig:runswvariance-b}\subref{fig:runswvariance-c}) which is lower for lower noise.  While the optimization eventually plateaus, it tracks the noiseless optimization until it approaches the plateaued energy.  This sensitivity to noise can be attributed to the non-linearity of the optimization.
We explicitly measure this non-linearity by looking at the distribution of one parameter after a noisy UBOS optimization (see Fig.~\ref{fig:param_opt}), finding that it is not centered at its true value;  this is to be contrasted with gradient descent, where the expectation value of the gradient over noisy estimates is correct.   In practice, this means that there may be a tradeoff between UBOS and SGD when expectation values are sufficiently noisy.   To further probe this, we look at the decrease in energy during different parts of the optimization (see Fig.~\ref{fig:onestep}).   At the beginning of the optimization, UBOS performs well even for the largest noise level studied, with energies similar to those found when optimizing with exact expectation values (which itself sometimes is slightly above the true global minimum due to imperfect classical optimization).  As the optimizations progresses, UBOS requires more precise estimates of matrix elements in order to continue to lower the energy.  Toward the end of the optimization, even low noise optimization often end up above the starting energies.  Finally, it is worth noting in practice that sometimes a small amount of noise in the estimates is actually beneficial in improving the energy reached by optimization (or reaching it faster).  For example, in the 8-site system, the smallest $\sigma$ runs reach lower energy than the exact run;  introducing slight noise to optimization has been shown to help in the related case of DMRG \cite{white_density_1992}.

\subsection{UBOS and the Barren Plateau}

\begin{figure*}[htb]%
\centering
\subfigure[][]{%
\label{fig:bpl-a}%
\includegraphics[height=140pt]{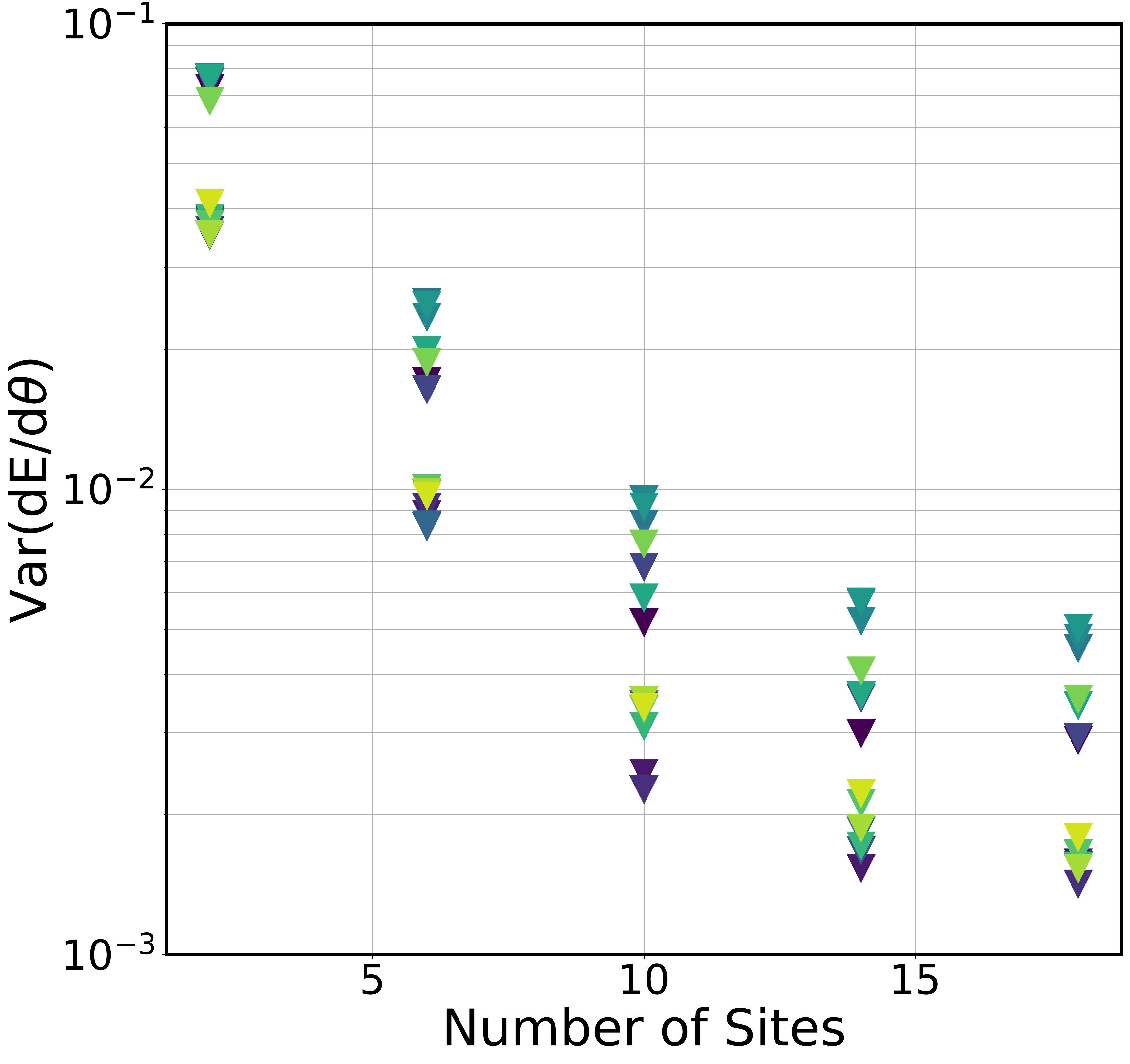}}%
\hspace{2pt}%
\subfigure[][]{%
\label{fig:bpl-b}%
\includegraphics[height=140pt]{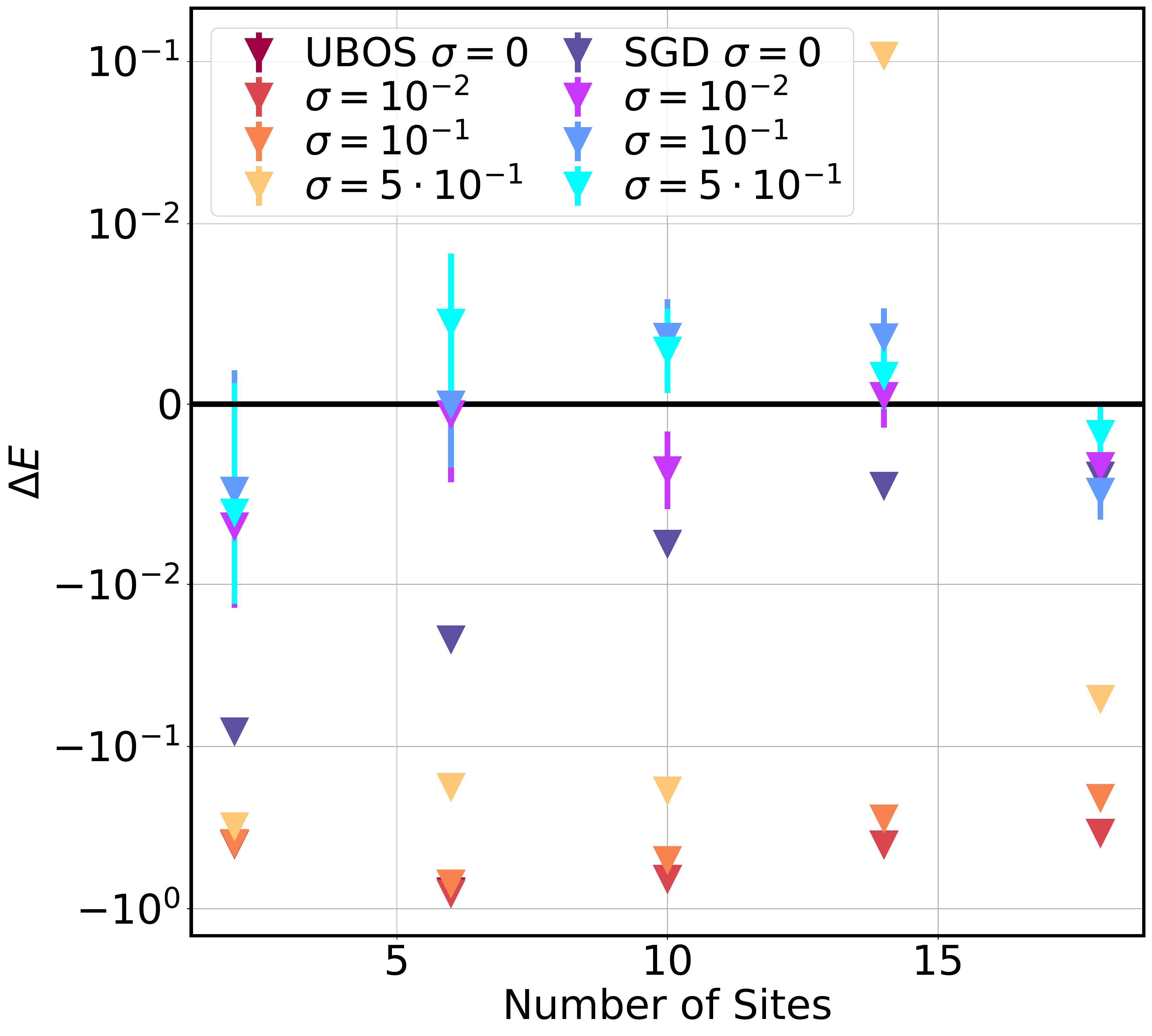}}%
\hspace{2pt}%
\subfigure[][]{%
\label{fig:bpl-c}%
\includegraphics[height=140pt]{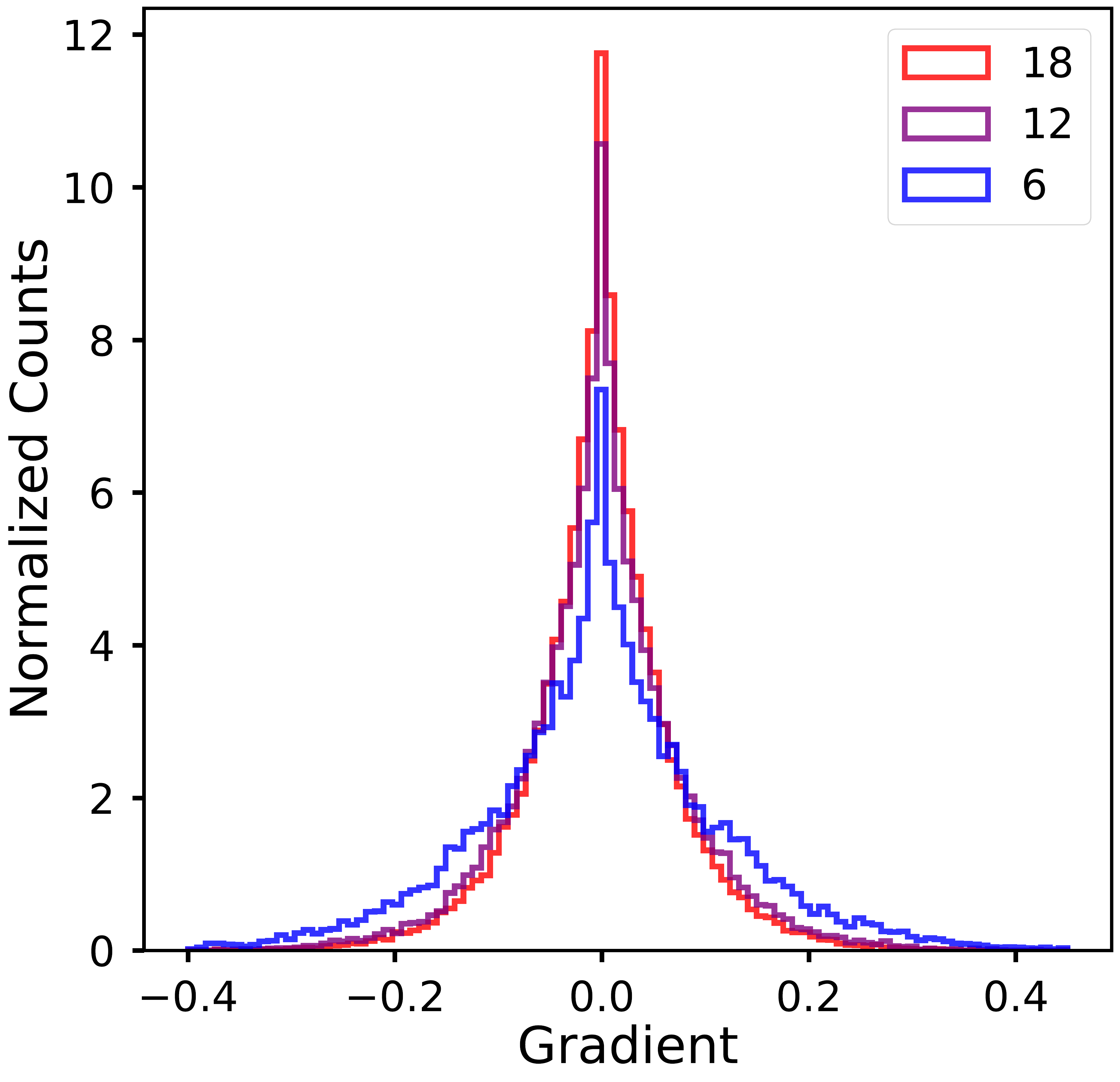}}%
\caption[]{Comparison between the average performance of SGD and UBOS for a gate in the middle of a 7 layer ansatz 
\subref{fig:bpl-a} Average initial variance of the gradient.  Each color represents one of the 15 parameters per gate. Error bars are smaller than the symbol size.
\subref{fig:bpl-b} Symmetric log plot of the average change in energy for a single SGD and UBOS step with and without Gaussian noise. 
\subref{fig:bpl-c} Normalized histograms of the gradients of randomly initialized ansatze with system sizes N = 6,12 and 18.
For both \subref{fig:bpl-a} and the SGD data in \subref{fig:bpl-b} averages are over 1000 randomly initialized ansatze. The UBOS data in \subref{fig:bpl-b} and the gradient histograms in \subref{fig:bpl-c} are averaged over 50 ansatze.}%
\label{fig:bpl}%
\end{figure*}

For a large class of random quantum circuits, the gradient of the objective function with respect to any parameter is expected to decrease exponentially with system size and number of gates \cite{mcclean_barren_2018}. Gradient descent performed on these circuits is difficult as even moderately noisy gradients make even the gradient direction unclear.
This is called the barren plateau problem. 
Here we give numerical evidence that UBOS is significantly less sensitive to barren plateaus. 
Starting from a randomly initialized circuit and picking a gate from the middle of the ansatz 
we compare the performance of UBOS and gradient descent. As expected from barren plateaus we see that the variance of the gradients decays exponentially with system size 
(out to a depth-dependent cutoff)
decreasing by an order of magnitude as we go from a 7 layer ansatz with 2 qubits to one with 18 qubits (see Fig.~\ref{fig:bpl-a}). The average gradient in our test for the 18 qubit ansatz has a magnitude of 0.026 to 0.046, depending on the parameter.  Over the same system sizes, the change in energy, $\Delta E$, due to a single (noiseless) UBOS step varies by at most ~50\%, with a typical change in energy of approximately $\Delta E \approx -0.6$. UBOS is still able to significantly lower the energy even when the gradients approach 0. 
Even more dramatically, for the gradients in this gate, only a modest amount of Gaussian noise is necessary
to obscure the sign of the gradient. In contrast, the effect of Gaussian noise on UBOS performance remains only weakly influenced by system size despite $|\Delta E|$ depending heavily on the amount of noise (see Fig.~\ref{fig:bpl-b}). This suggests that UBOS may be a useful tool for optimizing circuits in the barren plateau. 

\section{\label{sec:VITE} UBOS and Variational Imaginary Time Evolution}
In the previous sections, we have shown how UBOS is used in VQE to minimize the energy of variational ansatze. We now describe a time-evolution version of UBOS (TUBOS) with a focus here on variational imaginary time evolution (VITE); the generalization to real-time evolution is straightforward \cite{kochkov_variational_2018,mclachlan_time-dependent_1964,paeckel_time-evolution_2019,sorella_green_1998,schiro_time-dependent_2010}.  In fact, this discussion applies to any application where we wish to maximize an objective function $|\bra{\Phi(\boldsymbol{\theta})}\hat{O}\ket{\Psi(\boldsymbol{\theta}')}|$ with the variational parameters in one of the wavefunctions fixed. In the literature, there are several methods for performing VITE \cite{endo_variational_2020,mcardle_error-mitigated_2019,motta_determining_2020,yuan_theory_2019,barison_efficient_2021,benedetti_hardware-efficient_2020,lin_real-_2020}. For illustrative purposes we approximate the imaginary time evolution operator, $e^{-\beta \hat{H}}$, as $(1-\frac{\beta}{n} \hat{H})^n$. In order to approximate the action of $e^{-\beta \hat{H}}$ on a wavefunction, we then perform n steps where we variationally maximize $|\bra{\Psi(\boldsymbol{\theta}^{i+1})}1-\frac{\beta}{n}\hat{H}\ket{\Psi(\boldsymbol{\theta}^{i})}|$ in each step. During each step, the variational parameters for the wavefunction found in the previous step, $\boldsymbol{\theta}^{i}$, are kept fixed while $\boldsymbol{\theta}^{i+1}$ are not.

\begin{figure*}[htb]%
\centering
\subfigure[][]{%
\label{fig:tubos-a}%
\includegraphics[height=140pt]{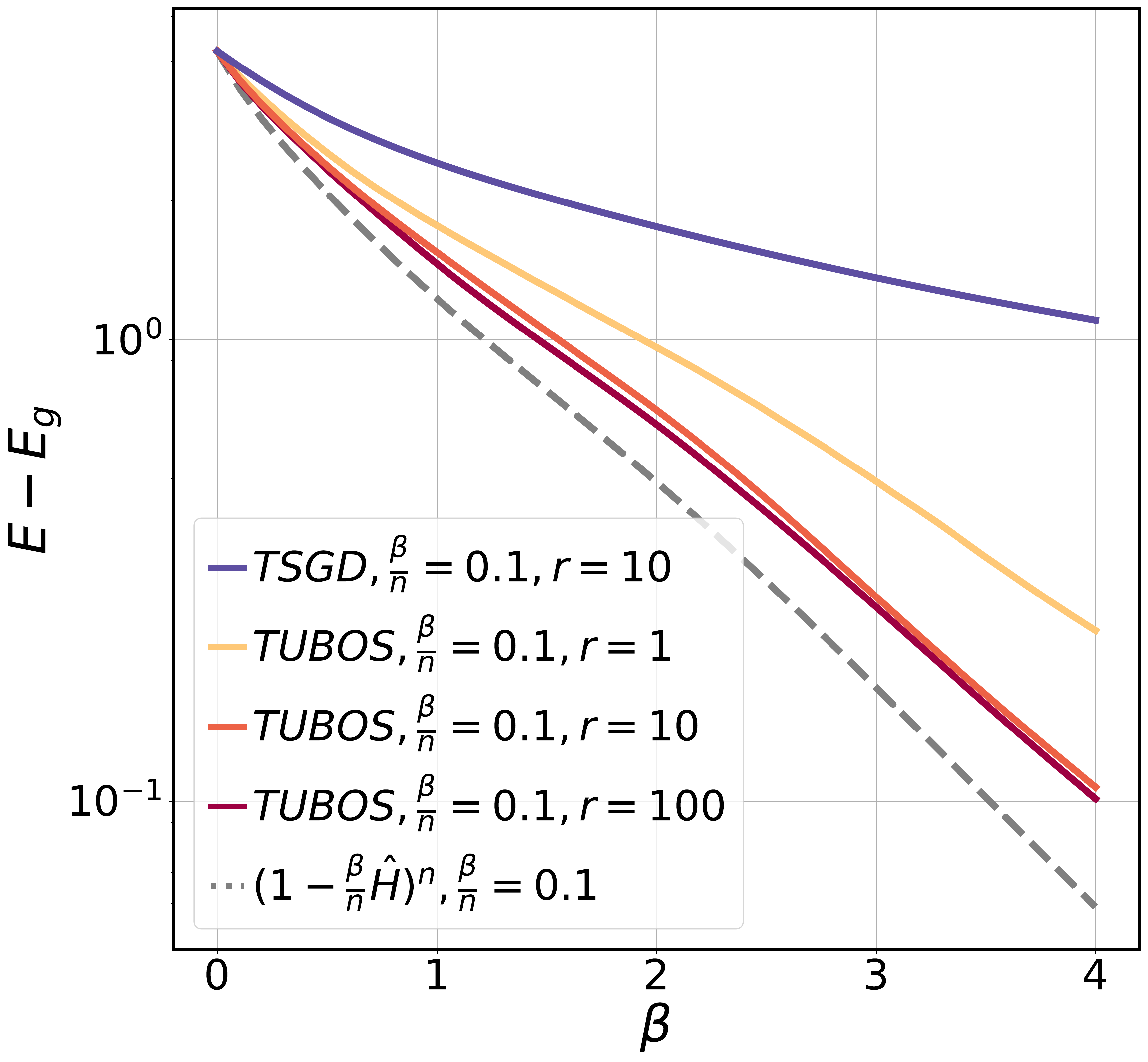}}%
\hspace{2pt}%
\subfigure[][]{%
\label{fig:tubos-b}%
\includegraphics[height=140pt]{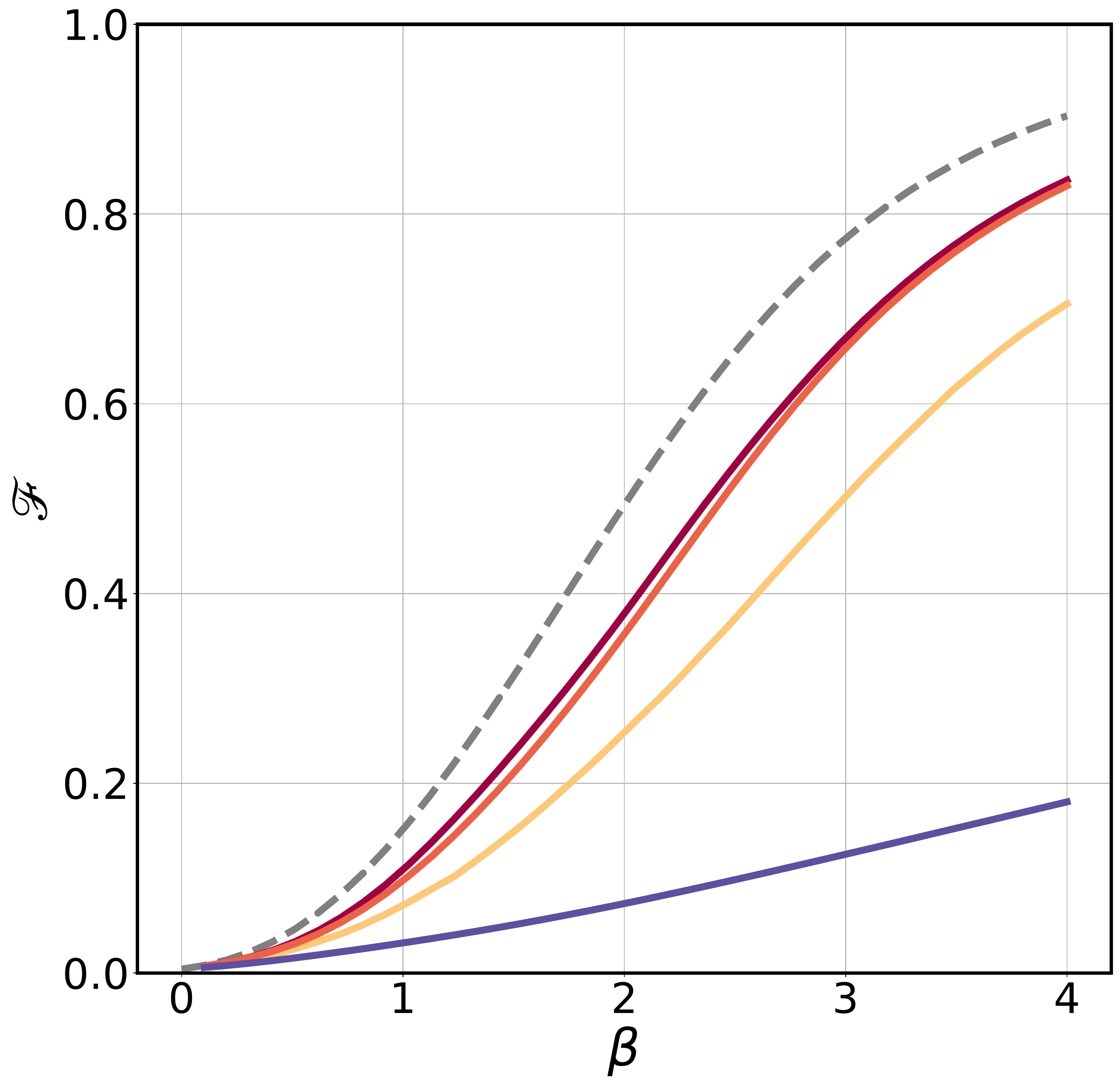}}%
\hspace{2pt}%
\subfigure[][]{%
\label{fig:tubos-c}%
\includegraphics[height=140pt]{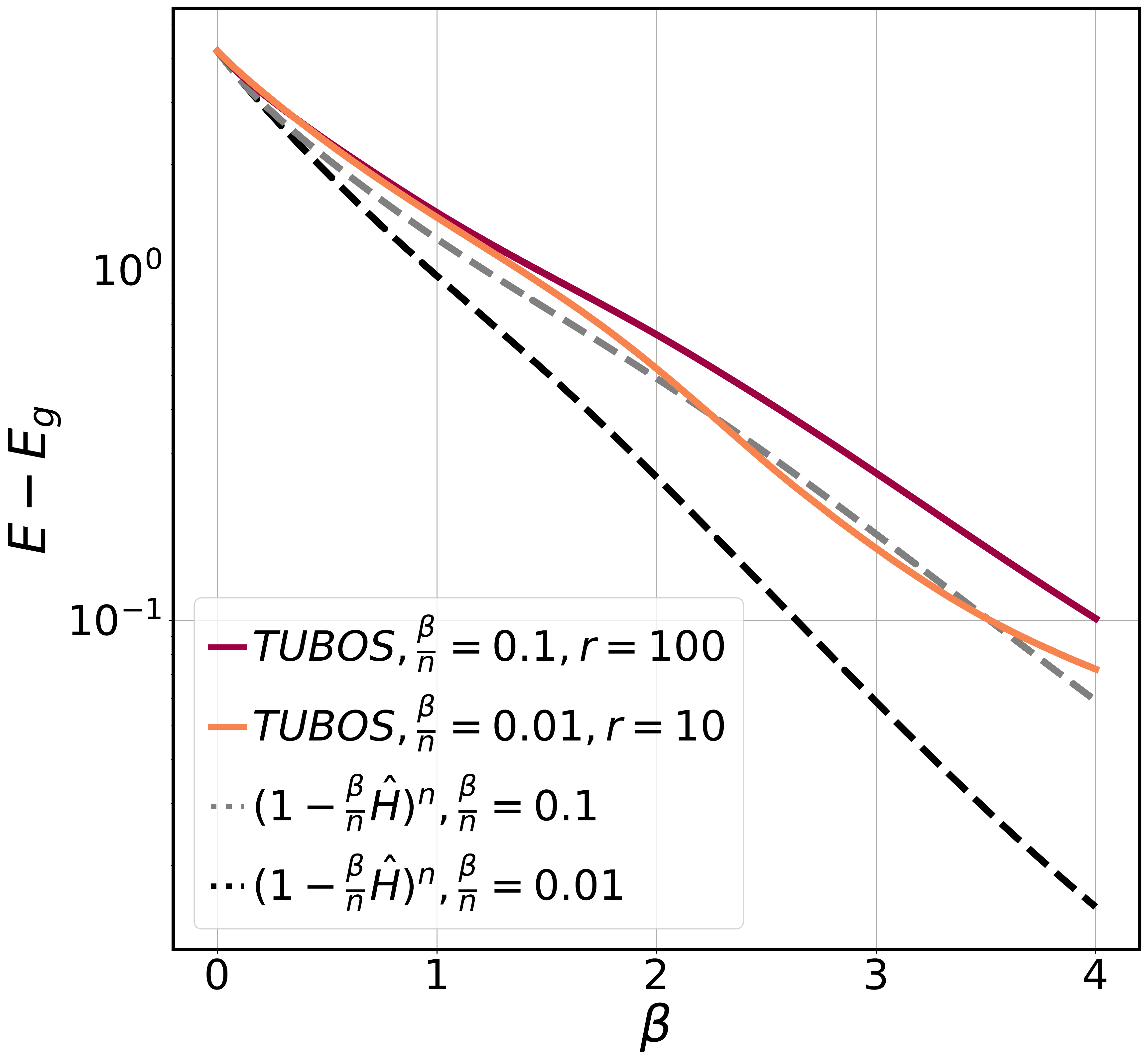}}%
\caption[]{We perform numerical experiments using TUBOS and its SGD counterpart (TSGD) on an 8 qubit 7 layer ansatz. Following the discussion in \ref{sec:VITE}, we simulate the action of $e^{-\beta \hat{H}}$ on our wavefunction. In \subref{fig:tubos-a} and \subref{fig:tubos-b}, we show noiseless simulations optimizing our wavefunction using TUBOS and TSGD with $\beta/n=0.1$. We sweep over each gate in the ansatz r times before we make a time step. The dashed lines represent the exact trajectories from applying $1-\frac{\beta}{n} \hat{H}$ to our ansatz. For fixed $\beta/n$ and r, TUBOS and TSGD require the same amount of expectation value measurements to perform but TUBOS drastically outperforms TSGD. In \subref{fig:tubos-c}, we compare results for TUBOS using $\beta/n=0.1$ and 0.01. Both TUBOS optimizations require the same amount of expectation values to perform but the smaller $\beta/n$ reaches lower energy.}%
\label{fig:tubos}%
\end{figure*}

Following the argument in Section \ref{sec:methods}, we can write
\begin{equation}
    \begin{split}
    &\bra{\Psi(\boldsymbol{\theta}^{i+1})}1-\frac{\beta}{n}\hat{H}\ket{\Psi(\boldsymbol{\theta}^{i})} \\ & = \sum_{\alpha\beta}t^{*\alpha\beta}_{j}\bra{\Psi(\boldsymbol{\theta}^{i+1})^{\alpha\beta}_{j}}1-\frac{\beta}{n}\hat{H}\ket{\Psi(\boldsymbol{\theta}^{i})}
    \end{split}
    \label{imag_time}
\end{equation}
where we have replaced the jth gate in the target wavefunction with a sum of two qubit Pauli matrices. To maximize the overlap we must measure each of the 16 summands, $\bra{\Psi(\boldsymbol{\theta}^{i+1})^{\alpha\beta}_{j}}1-\frac{\beta}{n}\hat{H}\ket{\Psi(\boldsymbol{\theta}^{i})}$. A quantum circuit to measure $\bra{\Psi(\boldsymbol{\theta}^{i+1})^{\alpha\beta}_{j}}\hat{O}\ket{\Psi(\boldsymbol{\theta}^{i})}$, for any unitary operator $\hat{O}$, can be constructed similarly to Fig. \ref{fig:gradmeasurecircuit} where each gate that is different between the initial and target state must be a controlled gate. We can then classically find the $\boldsymbol{t_j}$ that maximizes the overlap and perform an update to $\boldsymbol{\theta}^{i+1}$. We sweep over gates until a convergence criterion is met before moving to the i+1 step. If instead we wish to optimize Eq. \ref{imag_time} using SGD (TSGD), we would need to measure the same 16 elements. In Fig. \ref{fig:tubos}, we perform numerical experiments for TUBOS. TUBOS drastically out performs TSGD.

TUBOS differs from VQE UBOS in two ways.  First, the number of `matrix elements' needed is only 16 per two qubit gate. Secondly, each matrix element is somewhat more complicated to compute. One needs a controlled gate for ever gate that is different between $\ket{\Psi(\boldsymbol{\theta}^{i})}$ and the current $\ket{\Psi(\boldsymbol{\theta}^{i+1})}$; this is to be contrasted with the two control gates needed for UBOS. TUBOS has significant similarity to the Refs.~\onlinecite{lin_real-_2020} and \onlinecite{benedetti_hardware-efficient_2020} as well as the recently posted Ref. \onlinecite{barison_efficient_2021} which all perform either real or imaginary time evolution by iteratively maximizing the fidelity with a time-evolved state.  Such iterative fidelity maximization is also the key step in the imaginary time supervised wavefunction optimization (IT-SWO) approach to variational optimization \cite{kochkov_variational_2018}.  These four methods and TUBOS all are linear in the total number of ansatz parameters avoiding the quadratic cost in total parameters that plagues standard time dependent variational approaches such as stochastic reconfiguration. It is interesting to note that the number of expectation values required to reach a $E - E_g$ of $10^{-1}$ on the 8 qubit, 7 layer ansatz is similar for TUBOS with $\beta/n=0.1$, r=10 and UBOS.

These approaches to time evolution mainly differ in whether they optimize the maximum fidelity by SGD simultaneously on all parameters (Refs.~\onlinecite{kochkov_variational_2018} and \onlinecite{barison_efficient_2021}) versus subsets of the parameters at a time (TUBOS and Refs.~\onlinecite{lin_real-_2020},\onlinecite{benedetti_hardware-efficient_2020}) and whether each optimization step requires solving a non-trivial optimization problem (as in TUBOS) or taking a step which is either analytically solvable or proportional to the gradient.
It is also worth contrasting TUBOS and QITE \cite{motta_determining_2020}.  Both TUBOS and QITE generate an optimization problem over some gate.  TUBOS sweeps over fixed-size gates while QITE introduces, at each time step, a new  gate at the `top' of the circuit whose size is proportional to the time-step resulting in a cost which grows exponentially with depth, and reaching depths that might not be realizable on a NISQ device.   

\section{\label{sec:discussion} Discussion and Outlook}
In this paper, we introduced UBOS (and a time-evolved version TUBOS), a gradient-free and hyperparameter-free optimization method for the optimization of quantum circuits on hybrid variational algorithms such as VQE. UBOS works by iteratively sweeping over gates and generating an effective Hamiltonian $\tilde{H}$ (whose matrix elements are computed on a quantum computer). UBOS then classically finds the gate parameters that minimize the energy with respect to this effective Hamiltonian. We compared UBOS with standard stochastic gradient descent finding that, for the system sizes studied, UBOS consistently reached the same energy as stochastic gradient descent with an order of magnitude less expectation values.
UBOS is also able to tunnel through some local minima (i.e. those that exist for a single quantum gate). We observe, in situations where the gradient with respect to the parameters is extremely small, UBOS is still able to make non-trivial steps decreasing the energy; this is true even in the face of significant noise suggesting UBOS could also be a useful tool at the beginning of an optimization for escaping the barren plateau.

The effect on UBOS of stochastic noise on the matrix elements is more complicated than stochastic gradient descent which is unbiased in expectation.  In practice, the effect of this non-linearity results in a noise-dependent plateau of the UBOS energy above the minimum energy representable by the quantum circuit. This suggests an adaptive protocol where one uses progressively more measurements per expectation value as the optimization progresses. Alternatively, it may be advantageous to switch between UBOS and stochastic gradient descent as one gets closer to the minima.  One may also be able to correct this non-linearity with penalty method techniques \cite{ceperley_penalty_1999}.  

We have applied UBOS to a generic two-qubit unitary.  For devices which are not using this particular ansatz, it is straightforward to transfer such a pattern of gates to and from the specific architecture used in a machine \cite{tucci_introduction_2005} giving an optimization algorithm for the same quantum cost as standard UBOS.  This also applies to other general ansatz such as unitary coupled cluster, which can be decomposed as the application of (potentially non-local)  gates.
For certain types of gates (such as fSim, U(3) or CU(3) gates) even less Pauli terms need to be measured than in the generic case (see Appendix \ref{sec:AppA}).  At cost exponential in gate size, one can also use gates which span more than two qubits potentially reaching the ground state in less steps. While preliminary tests on the 1D XXZ Heisenberg model suggest this is not a advantageous in this case, it may be useful for other Hamiltonians.  The UBOS framework also gives an alternative to the various forms of variational real/imaginary time evolution and seems more efficient in expectation values and scaling with parameters then many alternative approaches. 
In summary, UBOS further improves our ability to use VQE to optimize quantum circuits.  Given that the primary difficult in successfully using VQE is optimizing a quantum circuit, UBOS gives an approach which is efficient and overcomes many of the technical issues with standard stochastic gradient descent.

\begin{acknowledgements}
 BKC acknowledges useful discussions with Aram Harrow and Edgar Solomonik.  We acknowledge support from the Department of Energy grant DOE de-sc0020165.
 This project is part of the Blue
Waters sustained-petascale computing project, which is supported by the National Science Foundation (awards OCI-0725070 and ACI-1238993) and the State of Illinois. Blue Waters is a joint effort of the University of Illinois at Urbana-Champaign and its National Center for Supercomputing Applications. 
This work also made use of the Illinois Campus Cluster, a computing resource that is operated by the Illinois Campus Cluster Program (ICCP) in conjunction with the National Center for Supercomputing Applications (NCSA) and which is supported by funds from the University of Illinois at Urbana-Champaign.
\end{acknowledgements}

\bibliographystyle{apsrev4-1}
\bibliography{bibliography}

\clearpage
\pagebreak
\appendix

\setcounter{section}{0}
\setcounter{equation}{0}
\counterwithout{equation}{section} 
\counterwithout*{equation}{section}
\renewcommand{\theequation}{A\arabic{equation}}
\renewcommand{\thesection}{\arabic{section}} 
\renewcommand\thefigure{A\arabic{figure}}  
\setcounter{figure}{0}   
\onecolumngrid

\section{\label{sec:AppA} UBOS Applied to Popular Gates}
Here we demonstrate how to extend UBOS to commonly used gates, which are particular cases of the generic two-qubit unitary treated in the main text.
\subsection{$U3$ Gate}
The one-qubit $U3$ gate
\begin{equation}
U3 (\theta, \lambda, \phi) =
    \begin{bmatrix}
    cos(\theta/2) & -e^{i\lambda}sin(\theta/2) \\
    e^{i\phi}sin(\theta/2) & e^{i(\lambda+\phi)}cos(\theta/2)
    \end{bmatrix}
\label{U3eq}
\end{equation}
is, up to a phase, able to represent any single-qubit unitary operation and requires all four single-qubit Pauli matrices in our representation.
\begin{equation}
    U_3=\sum_{\alpha=0}^{3}t^{\alpha}\sigma^{\alpha} \text{,}
\end{equation}
where $\sigma^0 = I$, $\sigma^1 = X$, $\sigma^2 = Y$, and $\sigma^3 = Z$.
The resulting subspace problem analogous to Eq. \eqref{EtHt} requires a 4x4 Hermitian matrix with 10 unique matrix elements.

\subsection{$CU3$}
Not all gates require all Pauli strings in order to construct them. Here we demonstrate that for a controlled one-qubit gate. The $CU3$ gate
\begin{equation}
CU3 (\theta, \lambda, \phi) =
  \begin{bmatrix}
    1 & 0 & 0 & 0 \\
    0 & 1 & 0 & 0 \\
    0 & 0 & cos(\theta/2) & -e^{i\lambda}sin(\theta/2) \\
    0 & 0 & e^{i\phi}sin(\theta/2) & e^{i(\lambda+\phi)}cos(\theta/2)
    
  \end{bmatrix}
\end{equation}
can be constructed from the 8 Pauli strings that share non-zero matrix elements with $CU3$:
\begin{equation}
\begin{split}
P_{00}=
 \begin{bmatrix}
    1\phantom{-} & 0\phantom{-} & 0\phantom{-} & 0\phantom{-} \\
    0\phantom{-} & 1\phantom{-} & 0\phantom{-} & 0\phantom{-} \\
    0\phantom{-} & 0\phantom{-} & 1\phantom{-} & 0\phantom{-} \\
    0\phantom{-} & 0\phantom{-} & 0\phantom{-} & 1\phantom{-}
  \end{bmatrix}
P_{01}=
 \begin{bmatrix}
    0\phantom{-} & 1\phantom{-} & 0\phantom{-} & 0\phantom{-} \\
    1\phantom{-} & 0\phantom{-} & 0\phantom{-} & 0\phantom{-} \\
    0\phantom{-} & 0\phantom{-} & 0\phantom{-} & 1\phantom{-} \\
    0 \phantom{-}& 0\phantom{-} & 1\phantom{-} & 0\phantom{-}
  \end{bmatrix} \\
P_{02}=
 \begin{bmatrix}
    0\phantom{-} & -i & 0\phantom{-} & 0\phantom{-} \\
    i\phantom{-} & 0\phantom{-} & 0\phantom{-} & 0\phantom{-} \\
    0\phantom{-} & 0\phantom{-} & 0\phantom{-} & -i \\
    0\phantom{-} & 0\phantom{-} & i\phantom{-} & 0\phantom{-}
  \end{bmatrix}
P_{03}=
 \begin{bmatrix}
    1\phantom{-} & 0\phantom{-} & 0\phantom{-} & 0\phantom{-} \\
    0\phantom{-} & -1 & 0\phantom{-} & 0\phantom{-} \\
    0\phantom{-} & 0\phantom{-} & 1\phantom{-} & 0\phantom{-} \\
    0\phantom{-} & 0\phantom{-} & 0\phantom{-} & -1
  \end{bmatrix} \\
P_{30}=
 \begin{bmatrix}
    1\phantom{-} & 0\phantom{-} & 0\phantom{-} & 0\phantom{-}\\
    0\phantom{-} & 1\phantom{-} & 0\phantom{-} & 0\phantom{-} \\
    0\phantom{-} & 0\phantom{-} & -1 & 0\phantom{-}\\
    0\phantom{-} & 0\phantom{-} & 0\phantom{-} & -1
  \end{bmatrix}
P_{31}=
 \begin{bmatrix}
    0\phantom{-} & 1\phantom{-} & 0\phantom{-} & 0\phantom{-} \\
    1\phantom{-} & 0\phantom{-} & 0\phantom{-} & 0\phantom{-} \\
    0\phantom{-} & 0\phantom{-} & 0\phantom{-} & -1 \\
    0\phantom{-} & 0\phantom{-} & -1 & 0\phantom{-}
  \end{bmatrix} \\
P_{32}=
 \begin{bmatrix}
    0\phantom{-} & -i & 0\phantom{-} & 0\phantom{-} \\
    i\phantom{-} & 0\phantom{-} & 0\phantom{-} & 0\phantom{-} \\
    0\phantom{-} & 0\phantom{-} & 0\phantom{-} & i\phantom{-} \\
    0\phantom{-} & 0\phantom{-} & -i & 0\phantom{-}
  \end{bmatrix}
P_{33}=
 \begin{bmatrix}
    1\phantom{-} & 0\phantom{-} & 0\phantom{-} & 0\phantom{-} \\
    0\phantom{-} & -1 & 0\phantom{-} & 0\phantom{-} \\
    0\phantom{-} & 0\phantom{-} & -1 & 0\phantom{-} \\
    0\phantom{-} & 0\phantom{-} & 0\phantom{-} & 1\phantom{-}
  \end{bmatrix}
\end{split}
\end{equation}
The resulting subspace problem analogous to Eq. \ref{EtHt} requires an 8x8 Hermitian matrix and the measurement of 40 unique matrix elements.

We could further constrain the subspace where $CU3$ gate is contained.
In particular, it is contained in the subspace of operators spanned by the following basis of operators:
\begin{align}
  R^0 &= CU3(0, 0, 0)=  \begin{bmatrix}
  1 & 0 & 0 & 0 \\
  0 & 1 & 0 & 0 \\
  0 & 0 & 1 & 0 \\
  0 & 0 & 0 & 1
  \end{bmatrix} \\
  R^1 &= CU3(0, \pi / 2, \pi / 2)=  \begin{bmatrix}
  1 & 0 & 0 & 0 \\
  0 & 1 & 0 & 0 \\
  0 & 0 & 1 & 0 \\
  0 & 0 & 0 & -1
  \end{bmatrix} \\
  R^2 &= CU3(2 \pi, \pi, 0)=  \begin{bmatrix}
  1 & 0 & 0 & 0 \\
  0 & 1 & 0 & 0 \\
  0 & 0 & 0 & 1 \\
  0 & 0 & 1 & 0
  \end{bmatrix} \\
  R^3 &= CU3(2 \pi, 0, 0) = \begin{bmatrix}
  1 & 0 & 0 & 0 \\
  0 & 1 & 0 & 0 \\
  0 & 0 & 0 & -1 \\
  0 & 0 & 1 & 0
  \end{bmatrix} \\
  R^4 &= CU3(2 \pi, 0, 0) = \begin{bmatrix}
  1 & 0 & 0 & 0 \\
  0 & 1 & 0 & 0 \\
  0 & 0 & -1 & 0 \\
  0 & 0 & 0 & -1
  \end{bmatrix} \text{,}
\end{align}
which would generate a $\tilde{H}$ matrix of size $5 \cross 5$, which requires 15 unique matrix elements, compared to 136 for a generic two qubit gate.

\subsection{U3 $\otimes$ U3}
A common, hardware-efficient gate set is a CNOT gate followed by a U3 gate on the control qubit and on the target qubit. In order to optimize this gate, we optimize U3 $\otimes$ U3 acting on the control and target qubit. For this gate, one must measure the full 16x16 Hermitian matrix $\hat{H}$. In addition to the unitarity constraint, one must also further constrain $\mathbf{t}$ so that the sum of the pauli strings remains equal to a U3 $\otimes$ U3 gate.

\subsection{fSim($\theta$, $\phi$)}
Another important gate in modern quantum computers is the fSim gate.
\begin{equation}
\label{fsim}
    {\rm fSim}(\theta,\phi) = \begin{bmatrix}
    1 & 0 & 0 & 0 \\
    0 & \cos{\theta} & -i\sin{\theta} & 0 \\
    0 & i\sin{\theta} & \cos{\theta} & 0 \\
    0 & 0 & 0 & e^{-i\phi}
    
  \end{bmatrix}
\end{equation}
Just like in the CU3 gate case, we notice that not all the two qubit Pauli strings have overlap with the fSim($\theta$,$\phi$) gate. Here we only need $P^{00},P^{03},P^{11},P^{12},P^{21},P^{22},P^{30}$ and $P^{33}$ to represent the gate in our sum of Pauli strings. Again, in order to optimize the gate, we need an 8x8 Hermitan matrix and 40 unique matrix elements.

We could further constrain the subspace where the fSim($\theta$, $\phi$) is contained.
In particular, it is contained in the subspace of operators spanned by the following basis of operators:
\begin{align}
  R^0 &= {\rm fSim}(0, 0)=  \begin{bmatrix}
  1 & 0 & 0 & 0 \\
  0 & 1 & 0 & 0 \\
  0 & 0 & 1 & 0 \\
  0 & 0 & 0 & 1
  \end{bmatrix} \\
  R^1 &= {\rm fSim}(0, \pi)=  \begin{bmatrix}
  1 & 0 & 0 & 0 \\
  0 & 1 & 0 & 0 \\
  0 & 0 & 1 & 0 \\
  0 & 0 & 0 & -1
  \end{bmatrix} \\
  R^2 &= {\rm fSim}(\pi, 0)=  \begin{bmatrix}
  1 & 0 & 0 & 0 \\
  0 & -1 & 0 & 0 \\
  0 & 0 & -1 & 0 \\
  0 & 0 & 0 & 1
  \end{bmatrix} \\
  R^3 &= {\rm fSim}(\pi / 2, 0) = \begin{bmatrix}
  1 & 0 & 0 & 0 \\
  0 & 0 & -i & 0 \\
  0 & i & 0 & 0 \\
  0 & 0 & 0 & 1
  \end{bmatrix} \text{,}
\end{align}
which would generate a $\tilde{H}$ matrix of size $4 \cross 4$, which requires 10 unique matrix elements, compared to 136 for a generic two qubit gate.

We note that many gates may be  able to be represented with fewer linearly independent unitary matrices than the naive use of Pauli matrices as we have demonstrated for $CU3$ and fSim.

\section{\label{sec:AppB} Quantum Circuits Used in Simulations}
Here we demonstrate the approach and quantum circuits for measuring the ansatz gradients, $\frac{dE}{d\boldsymbol{\theta}}=\frac{\bra{\Psi(\boldsymbol{\theta})}\hat{H}\ket{\Psi(\boldsymbol{\theta})}}{d\boldsymbol{\theta}}$, and the matrix elements $\tilde{H}^{\alpha\beta\alpha'\beta'}=\bra{\Psi(\boldsymbol{\theta})_{j}^{\alpha'\beta'}}\hat{H}\ket{\Psi(\boldsymbol{\theta})_{j}^{\alpha\beta}}$ where $\ket{\Psi(\boldsymbol{\theta})_{j}^{\alpha\beta}}$ is the ansatz $\ket{\Psi(\boldsymbol{\theta})}$ with the jth gate replaced by $P_{j}^{\alpha\beta}$. We have included the implicit reference to the circuit parameters, $\boldsymbol{\theta}$, in our discussion for this section.
We note that while we show how to use a Hadamard test circuit to measure expectation values of local operators there are other measurement schemes in the literature \cite{mitarai_methodology_2019,guerreschi_practical_2017}.

\subsection{Analytic Gradients for Generic Unitary Circuit Ansatz}

Using the Cartan (KAK) decomposition for $U \in SU(4)$ \cite{tucci_introduction_2005}, we can write our two qubit, 15 real parameter unitary operator as:
\begin{equation}
    U=(A_0 \otimes A_1)(e^{-i\vec{k} \cdot\vec{\Sigma}})(B_0 \otimes B_1)
\end{equation}
where $A_0$, $A_1$, $B_0$ and $B_1$ $\in SU(2)$ (in our case parameterized as Eq. \eqref{U3eq}), $\vec{k} \in \mathbb{R}^{3}$ and $\vec{\Sigma} = (P^{00},P^{11},P^{22})$. Using the KAK decomposition, we find an analytical expression for Eq.~\ref{paulidecomp} where the summation coefficients, $t_j^{\alpha\beta}(\boldsymbol{\theta}_j)$, are functions of the 15 circuit parameters specifying the jth gate. We use Python's SymPy package to find analytical expressions for the derivative of $U_j$ with respect to $\boldsymbol{\theta}_j$ as:
\begin{equation}
\frac{dU}{d\boldsymbol{\theta}_j}=\sum_{\alpha,\beta=0}^{3}{\frac{dt_j^{\alpha\beta}(\boldsymbol{\theta}_j)}{d\boldsymbol{\theta}_j}P^{\alpha\beta}}
\end{equation}

\begin{figure}[htp]
\begin{tabular}{c}
    \Qcircuit @C=1.4em @R=1.4em {
        \lstick{\ket{0}} & \multigate{1}{U_0} & \qw & \multigate{1}{U_5} & \qw & \qw & \qw & \multigate{1}{U_5^\dag} & \qw & \qw \\
        \lstick{\ket{0}} & \ghost{U_0} & \multigate{1}{U_3} & \ghost{U_5} & \multigate{1}{U_8} & \qw  & \multigate{1}{U_8^\dag} & \ghost{U_5^\dag} & \qw & \qw \\
        \lstick{\ket{0}} & \multigate{1}{U_1} & \ghost{U_3} & \multigate{1}{U_6} & \ghost{U_8} & \multigate{1}{\hat{O}} & \ghost{U_8^\dag} & \multigate{1}{U_6^\dag} &\qw & \qw \\
        \lstick{\ket{0}} & \ghost{U_1} & \multigate{1}{P_4^{\alpha\beta}} & \ghost{U_5} & \multigate{1}{U_{9}} & \ghost{\hat{O}} & \multigate{1}{U_9^\dag} & \ghost{U_6^\dag} & \multigate{1}{U_4^\dag/P_4^{\alpha'\beta'}} & \qw\\
        \lstick{\ket{0}} & \multigate{1}{U_2} & \ghost{P_4^{\alpha\beta}} & \multigate{1}{U_7} & \ghost{U_{9}} & \qw & \ghost{U_9^\dag} & \multigate{1}{U_7^\dag} & \ghost{U_4^\dag/P_4^{\alpha'\beta'}} & \qw \\
        \lstick{\ket{0}} & \ghost{U_2} & \qw & \ghost{U_7} & \qw & \qw & \qw & \ghost{U_7^\dag} & \qw & \qw \\
        \lstick{\ket{0}} & \gate{H} & \ctrl{-2} & \qw & \qw & \ctrl{-3} & \qw & \qw & \ctrl{-2} & \gate{S^b} & \gate{H} & \qw & \expval{Z}
        }
\end{tabular}
    \caption{Quantum circuit for the measurement of an operator gradient, $\mel{\Psi(\boldsymbol{\theta})}{\hat{O}}{\frac{\Psi(\boldsymbol{\theta})^{\alpha\beta}}{d\boldsymbol{\theta}_j}}$, and an operator subspace matrix element $\mel{\Psi(\boldsymbol{\theta})_{j}^{\alpha'\beta'}}{\hat{O}}{\Psi(\boldsymbol{\theta})_{j}^{\alpha\beta}}$. The controlled $U_4$ gate is used when measuring the gradient while the controlled $P^{\alpha'\beta'}$ is used for the matrix elements. The measurement circuit shown is an example for j = 4 and a 6 qubit, 4 layer ansatz. When b = 0 (b=1), the real (imaginary) part of the expectation values are estimated by $-\expval{Z}$.}
\label{fig:gradmeasurecircuit}
\end{figure}
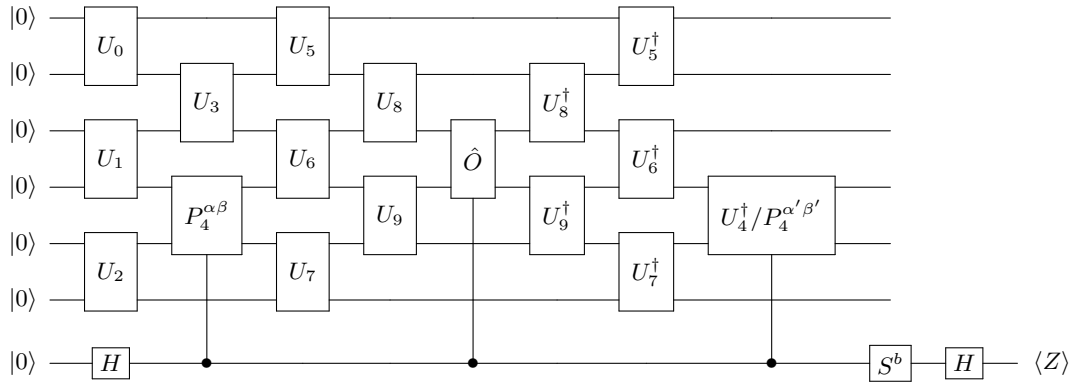

The expression for the gradients of the jth gate then becomes
\begin{equation}
\begin{split}
    \frac{dE}{d\boldsymbol{\theta}_j}&=\frac{\bra{\Psi(\boldsymbol{\theta})}\hat{H}\ket{\Psi(\boldsymbol{\theta})}}{d\boldsymbol{\theta}} \\
     &=2\Re\left(\mel{\Psi(\boldsymbol{\theta})}{\hat{H}}{\frac{\Psi(\boldsymbol{\theta})}{d\boldsymbol{\theta}_j}} \right) \\
     &=2\Re\left(\sum_{\alpha,\beta=0}^{3}\frac{dt_j^{\alpha\beta}(\boldsymbol{\theta}_j)}{d\boldsymbol{\theta}_j}\mel{\Psi(\boldsymbol{\theta})}{\hat{H}}{\Psi(\boldsymbol{\theta})_{j}^{\alpha\beta}} \right)
\end{split}
\label{gradequation}
\end{equation}
By measuring the 16 expectation values in Eq. \eqref{gradequation}, one can compute the gradients jth gate. In Fig.~\ref{fig:gradmeasurecircuit} we show how to measure the expectation values with an example Hadamard test circuit.

\subsection{Matrix Elements and Quantum Circuits for UBOS}

For UBOS, we must compute the matrix elements of $\tilde{H}$ (Eq. \eqref{Htilde}). For an arbitrary unitary $U \in SU(4)$, we measure the 136 unique elements, $\Tilde{H}^{\alpha'\beta'\alpha\beta} = \mel{\Psi(\boldsymbol{\theta})_{j}^{\alpha'\beta'}}{\hat{H}}{\Psi(\boldsymbol{\theta})_{j}^{\alpha\beta}}$. In Fig. \ref{fig:gradmeasurecircuit} we show to to measure these expectation values with an example Hadamard test circuit.

\section{\label{sec:AppC} Counting Expectation Values Needed}

\begin{figure}
    \centering
    \includegraphics[width=0.8\linewidth]{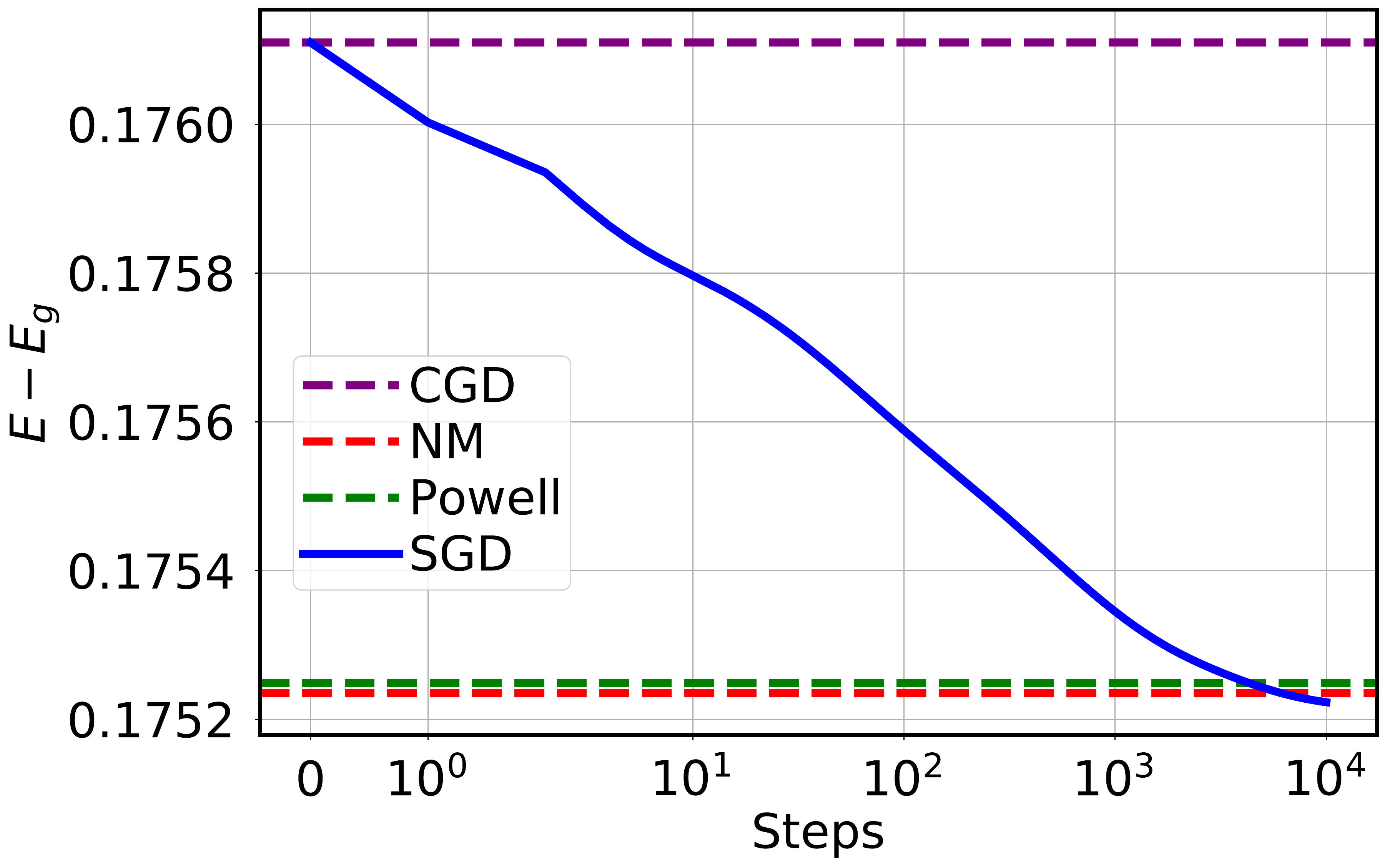}
    \caption{Single gate optimization comparison. After 100 UBOS gate optimizations on a 12 site 7 layer ansatz for the 1D XXZ model. We show the results for the lowest energy found by 4 different optimizers (conjugate gradient descent (CGD), Nelder-Mead (NM), Powell and SGD) acting on a single gate using UBOS. We note that SGD can either be implemented using quantum measurements of the gradients or using UBOS with $\hat{H}$. For our UBOS numerical experiments, we use NM as our classical optimizer because it regularly performed better than CGD and Powell. One could also choose to do SGD using $\hat{H}$. As in the figure, this occasionally outperforms NM but usually at a higher classical computational cost.}
    \label{fig:class_opt}
\end{figure}

We calculated the number of measurement circuit expectation values needed, $N_{evn}$, for an optimization run as
\begin{equation}
    N_{evn}=N_{epochs} \times N_{\hat{H}_{meas}} \times N_{operators}
\end{equation}
where $N_{epochs}$ is the number of training epochs, $N_{\hat{H}_{meas}}$ is the number of matrix elements/gradients with the Hamiltonian and $N_{operators}$ is the number of unique local operators in the Hamiltonian. We note that different strategies for grouping Hamiltonian operators can reduce the number of distinct circuits needed to calculate Hamiltonain expectation values \cite{kokail_self-verifying_2019,fizmaylov_revising_2019,gokhale_on3_2020}.

\section{\label{sec:AppD} Single Gate Optimizer Comparison}

In order to better understand the potential advantage offered by UBOS, we used UBOS to optimize a 12 site 7 layer ansatz for our 1D XXZ model. In the midst of the optimization, we then performed SGD on a single gate and compared with a single UBOS step using different classical optimizers to solve Eq. \ref{EtHt} (Fig. \ref{fig:class_opt}). We find in this example (and others not shown) that classically optimizing with both SGD and Nelder-Mead  outperforms other classical optimization techniques. In practice, we find SGD reaches lower in energy at significant classical computation cost. It is an interesting open question to further explore classical optimization methods to determine better approaches to optimizing for UBOS.  Notice that despite the fact that SGD was best classically, it is still superior to do it classically in this UBOS context instead of doing quantum SGD since it requires significantly fewer measurements. 
When comparing against doing SGD with all the gates, we find only two orders of magnitude difference between the number of steps required to converge to similar energy (see Fig. \ref{fig:14n7layers}) compared to the three orders of magnitude we see for a single gate.  Doing SGD with all the gates is more efficient than a single gate at a time because there are marginal returns toward the end of a single-gate optimization.  

\begin{figure}
    \centering
    \includegraphics[width=0.8\linewidth]{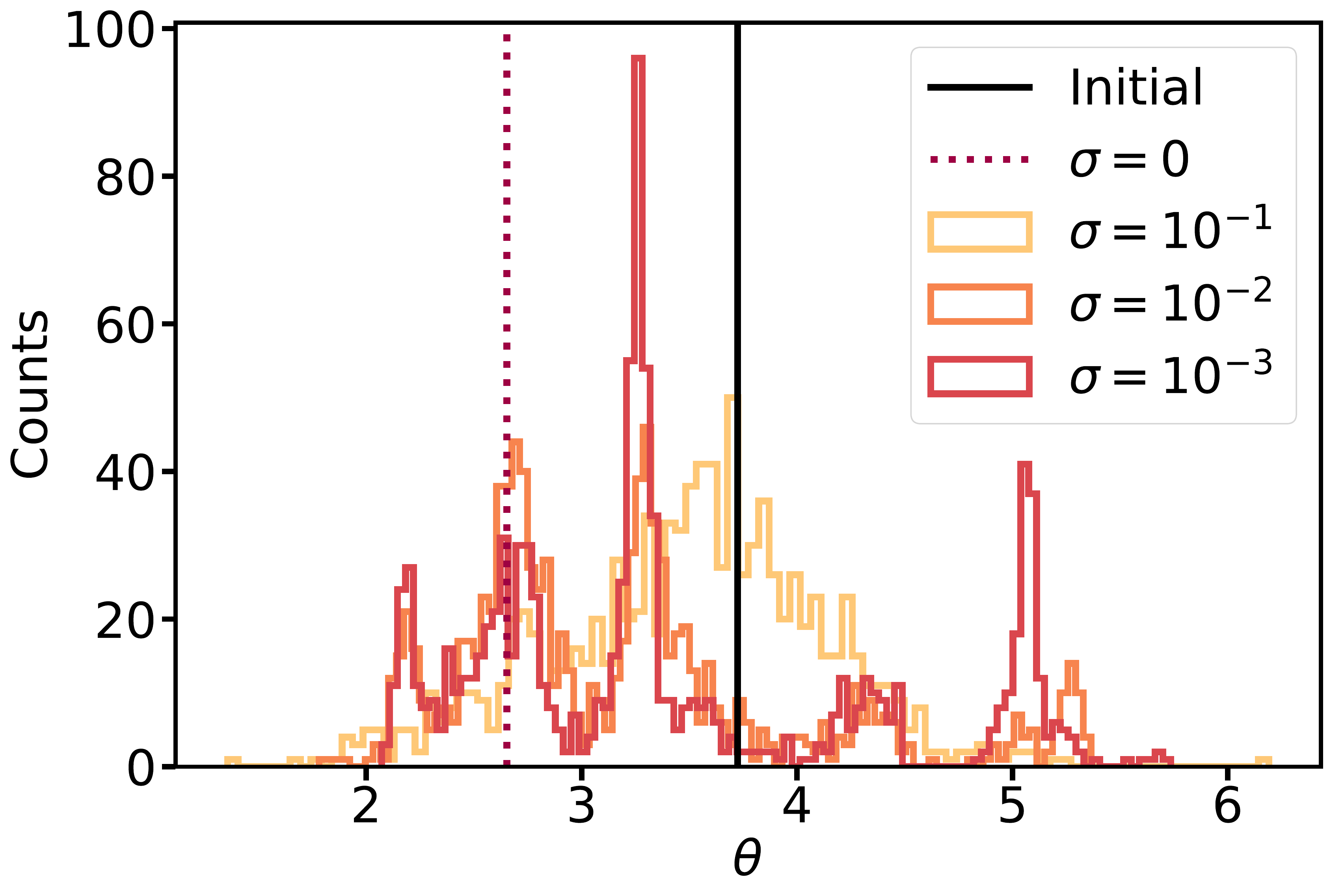}
    \caption{Plot for an optimized parameter of a gate after an UBOS optimization step. We randomly initialize a 12 site 7 layer ansatz and perform a single UBOS step for the XXZ 1D model. For the same randomly initialized ansatz, we optimize 1000 instances of $\tilde{H}$ with added Gaussian noise, $\sigma$, added to the matrix elements.}
    \label{fig:param_opt}
\end{figure}

Because of the unitary constraints imposed on our variational ansatz, finding the parameters that minimize Eq.~\ref{EtHt} is a nonlinear problem.  We find that the choice of classical optimizer can affect the minima found (Fig.~\ref{fig:class_opt}). We also find that our classical optimizers often have trouble finding the exact minima and that adding a small amount of Gaussian noise to the matrix elements of $\tilde{H}$ can sometimes result in our optimizer finding a lower energy solution (Fig.~\ref{fig:onestep}), especially at the beginning of optimization. In Fig~\ref{fig:param_opt}, we examine the distribution of an optimized gate parameter when noise has been added to the matrix elements of $\tilde{H}$. The distribution of this single parameter is multimodal in the presence of a small amount of noise indicating that the optimization problem is nonlinear. Following this set of non-linear steps appears to eventually reach a local minimum in this space. 

\section{Generalization of UBOS: Partially Optimizing Many Gates at Once}
\label{sec:large_unitaries}

In this work, we have applied UBOS to the subspace spanned by linearly independent matrices that fully contains the manifold defining a particular parameterized gate (such as $U4$, fSim, or $CU3$ gates).
In this section we show how UBOS can be generalized to \emph{partially} optimize large unitaries which can include a large number of gates and a large number of qubits, at a much lower cost than the naive $4^{\# {\rm qubits}}$ scaling.
While we have not explored this method in the present work, it would be interesting to study it in future works, since it could have potential advantages avoiding local minima and other difficult optimization landscapes.

\subsection{Algorithm}

Let us have a parameterized unitary operator defined over a Hilbert space of dimension $2^n$, $U(\theta_1, \ldots, \theta_p)$, where $\{ \theta_i\}$ are real parameters.
Similar to Eq.~\eqref{paulidecomp}, $U$ can be decomposed as:
\begin{align}
\label{large_U_decomp}
U(\theta_1, \ldots, \theta_p) = \sum_{\vec{\alpha}} r^{\alpha}(\theta_1, \ldots, \theta_p) R^{\alpha}
\end{align}
where $R^{\alpha}$ is a basis of operators.
In most of the present work, the elements in this basis are Pauli strings, but this is not a necessary condition for UBOS to work.
While applying UBOS directly to this unitary requires building a $\tilde{H}$ matrix of size $4^n \cross 4^n$, we can instead focus on a subspace $V$ spanned by a small subset of Pauli strings, $S$: $V = {\rm span}\{ R^{\alpha} : R^{\alpha} \in S \}$.
In general, subspace $V$ may only contain a sub-manifold of $U(\theta_1, \ldots, \theta_p)$, $\tilde{U}(\phi_1, \ldots, \phi_q)$, with $q \leq p$.
We can then optimize all parameters $\{ \phi_i \}$ at once after building a matrix $\tilde{H}$ of size $|S| \times |S|$.

In practice, unitary $U$ can be comprised of a large set of gates in our parameterized quantum circuit.
These gates can act on many qubits or could even be applied repeatedly over a set of qubits at different stages of the circuit.
The manifold defined by $U(\theta_1, \ldots, \theta_p)$ is therefore the tensor product of the manifolds defined by each parameterized gate.

One should be careful when choosing the optimization subspace, $V$, for two reasons.
First, it is not guaranteed that the intersection of manifold $U$ with $V$ is not empty.
In general, it is advantageous to choose a small dimensional space $V$ that contains a manifold with as many independent parameters as possible.
Second, one should keep in mind if it is desired to have $V$ contain the starting point of the optimization, \emph{i.e.}, $U(\bar{\theta_1}, \ldots, \bar{\theta_p})$, where $\{\bar{\theta_1}, \ldots, \bar{\theta_p}\}$ is the set of parameters found in the previous iteration.
If that is the case, it is easy to solve this problem by just including $U(\bar{\theta_1}, \ldots, \bar{\theta_p})$ as one of the vectors in the basis. 
Note that this does not guarantee that the intersection of this manifold with $V$ is a sub-manifold with dimension larger than zero.

\subsection{Example}
For illustration purposes, let us work out a concrete example of this general procedure.
Let $U(\theta_1, \theta_2, \phi_1, \phi_2)$ be the tensor product of two fSim gates of Eq.~\eqref{fsim} over four qubits: one over qubits labelled 1 and 2, and the other over qubits labelled 3 and 4.
In general, the optimization of a four qubit unitary would imply the construction of a $\tilde{H}$ of size $4^8 \times 4^8$.
However, if we focus on a dimension 4 subspace $V$ spanned by the 4 tensor products of operators $A$ and $B$:

\begin{align}
  A_j &= {\rm fSim}(\bar{\theta_j},\bar{\phi_j}) = \begin{bmatrix}
  1 & 0 & 0 & 0 \\
  0 & \cos{\bar{\theta_j}} & -i\sin{\bar{\theta_j}} & 0 \\
  0 & i\sin{\bar{\theta_j}} & \cos{\bar{\theta_j}} & 0 \\
  0 & 0 & 0 & 1
  \end{bmatrix} \\
  B_j &= {\rm fSim}(\bar{\theta_j},\bar{\phi_j}) = \begin{bmatrix}
  1 & 0 & 0 & 0 \\
  0 & \cos{\bar{\theta_j}} & -i\sin{\bar{\theta_j}} & 0 \\
  0 & i\sin{\bar{\theta_j}} & \cos{\bar{\theta_j}} & 0 \\
  0 & 0 & 0 & -1
  \end{bmatrix} \text{,}
\end{align}

\emph{i.e.}, if we have a basis of operators $S = \{ R^0 \equiv A_{1, 2} \otimes A_{3, 4}, R^1 \equiv A_{1, 2} \otimes B_{3, 4}, R^2 \equiv B_{3, 4} \otimes A_{1, 2}, R^3 \equiv B_{1, 2} \otimes B_{3, 4} \}$, where each operator $A$ or $B$ is applied to either the tuple of qubits $\{ 1, 2 \}$ or $\{ 3, 4 \}$, then we can optimize angles $\phi_{1, 2}$ and $\phi_{3, 4}$ simultaneously by building a matrix of size $4 \cross 4$, with only 10 independent entries.
An optimization in this subspace allows us to optimize $\phi_{1, 2}$ and $\phi_{3, 4}$ while keeping $\theta_{1, 2}$ and $\theta_{3, 4}$ fixed.
In this example, the coefficients $\{r^i\}$ are given by:
\begin{align}
\label{solution_example}
    r^0 &= \left(\frac{1 + e^{-i\phi_{1, 2}}}{2}\right) \left( \frac{1 + e^{-i \phi_{3, 4}}}{2}\right) \\
    r^1 &= \left(\frac{1 + e^{-i\phi_{1, 2}}}{2}\right) \left( \frac{1 - e^{-i \phi_{3, 4}}}{2}\right) \\
    r^2 &= \left(\frac{1 - e^{-i\phi_{1, 2}}}{2}\right) \left( \frac{1 + e^{-i \phi_{3, 4}}}{2}\right) \\
    r^3 &= \left(\frac{1 - e^{-i\phi_{1, 2}}}{2}\right) \left( \frac{1 - e^{-i \phi_{3, 4}}}{2}\right)
\end{align}
and the subspace $V$ includes the starting point $(\bar{\theta}_{1, 2}, \bar{\theta}_{3, 4}, \bar{\phi}_{1, 2}, \bar{\phi}_{3, 4})$, which is given by evaluating Eqs.~\eqref{solution_example} at $(\bar{\phi}_{1, 2}, \bar{\phi}_{3, 4})$.

\subsection{When is this procedure advantageous?}

While applying this procedure to the optimization of a parametrized circuit is more complicated than the UBOS procedure explored in this paper, its application could provide some benefits.
In the case that the simpler version of UBOS gets stuck in a local minimum, switching momentarily to this procedure, in order to optimize different sets of parameters at once, possibly comprising a set of gates defined over a larger number of qubits, could help get out of this minimum.
Furthermore, the flexibility in choosing subspace $V$ lets us try to find sub-manifolds of $U$ that, while have many free parameters, are contained in as low dimensional a subspace as possible.
This can give rise to more frugal optimization procedures than the naive parameterization of gates imposes.

\section{UBOS on a Manifold of Arbitrary Parameters}

Here we describe another viewpoint for using an UBOS-like algorithm to optimize an arbitrary subset of parameters  ${\Phi} \equiv \{\phi_1,...\phi_r\}$ which specify the wave-function generated by a quantum circuit $\ket{\Psi({\Phi)}}$.  We use as a basis for this manifold $k$ randomly selected wave-functions $\ket{\Psi({\Phi_1)}},\ket{\Psi({\Phi_2)}},...,\ket{\Psi({\Phi_k)}}$ where $k$ is chosen to be sufficiently large that the basis spans the full manifold.  Generic points from this manifold are likely to be linearly independent and so will form a (non-orthogonal) basis.  To determine $k$ we simply keep adding states to our basis until their rank saturates.  This rank can be determined in various ways: (1) evaluating it classically by expanding each $\ket{\Psi({\Phi_i)}}$ into a Pauli basis over the gates which are spanned by the parameters $\Phi_i$;  (2) compute the rank of the overlap matrix $S$ whose matrix elements $S_{ij}=\braket{\Psi({\Phi_i)}}{\Psi({\Phi_j)}}$ can be evaluated by quantum circuits.

Given the basis, as in UBOS, we  generate the effective Hamiltonian $\tilde{H}$ in that basis evaluating each matrix element  $\tilde{H_{ij}} = \bra{\Psi({\Phi_i)}}\hat{H}\ket{\Psi({\Phi_j)}}$ with a quantum circuit.  Alternatively, we can compute (enough of) $\tilde{H}$ by using the least-squares approach described in \ref{sec:HfromEs} on random states  in our manifold.  In this latter case, one can just choose random states until $\tilde{H}$ stabilizes and not separately generate a basis of full rank. 

Given the effective Hamiltonian $\tilde{H}$ in our basis, one can optimize as in standard UBOS by classically searching over random states in the $\Phi$ manifold and computing and optimizing their energy using $\tilde{H}$.  The only subtlety is one needs to evaluate the overlap between the basis elements and arbitrary states in the manifold $\braket{\Psi({\Phi_i)}}{\Psi({\Phi)}}$ for arbitrary $\Phi$ to accomplish this.  This can be done but might (in specific scenarios) involve large classical cost.  (While this could be done at reasonable quantum cost, since it needs to be done at each step of the optimization this would result in a process that's much closer to SGD where the full energy landscape of a subset of parameters couldn't be evaluated classically without constant re-evaluation on the quantum computer). 

\section{\label{sec:HfromEs} Finding $\tilde{H}$ from a System of Equations}

Previously we demonstrated how to find the effective Hamiltonian, $\tilde{H}$ by measuring each matrix element of $\tilde{H}$ with a quantum circuit of depth at most $2d$. Here we describe an alternative approach which uses only circuits of depth $d$, but whose resilience to stochastic noise is to be explored in future work. The high level ideas of this approach is as follows. In Eq.~\ref{EtHt} of the main text, the matrix elements in $\tilde{H}$ are linear unknowns.

Therefore, one can measure the energy, $E(\theta_1, \ldots, \theta_p)$, for different, randomly chosen $(\theta_1, \ldots, \theta_p)$ and construct a linear system of equations that one can use to solve for the matrix elements in $\tilde{H}$.  In the rest of the section we more explicitly describe this approach. 

Note that, similar to Eq.~\ref{EtHt} of the main text, the energy functional can always be written as a quadratic form in the entries of the vector $\mathbf{t_j}(\theta_1, \ldots, \theta_p)$.
In particular:
\begin{align}
\label{quadratic_E}
    E(\theta_1, \ldots, \theta_p)
    = &\sum_{\alpha, \alpha^\prime} t_j^\alpha(\theta_1, \ldots, \theta_p) \tilde{H}^{\alpha \alpha^\prime} t_j^{\alpha^\prime}(\theta_1, \ldots, \theta_p) \nonumber \\
    = &\sum_\alpha t_j^\alpha(\theta_1, \ldots, \theta_p) t_j^\alpha(\theta_1, \ldots, \theta_p)^* \tilde{H}^{\alpha \alpha} \nonumber \\
    &+ \sum_{\alpha < \alpha^\prime} \left\{ {\rm Re}[t_j^\alpha(\theta_1, \ldots, \theta_p) t_j^{\alpha^\prime}(\theta_1, \ldots, \theta_p)^*] {\rm Re}[\tilde{H}^{\alpha \alpha^\prime}] \right. \nonumber \\
    &\left.- {\rm Im}[t_j^\alpha(\theta_1, \ldots, \theta_p) t_j^{\alpha^\prime}(\theta_1, \ldots, \theta_p)^*] {\rm Im}[\tilde{H}^{\alpha \alpha^\prime}] \right\} \text{,}
\end{align}
where, due to the hermiticity of $\tilde{H}$, $\tilde{H}^{\alpha \alpha}$ is a real number .
We now define $t_{j, R}^{\alpha \alpha^\prime}(\theta_1, \ldots, \theta_p) \equiv {\rm Re}[t_j^\alpha(\theta_1, \ldots, \theta_p)t_j^{\alpha^\prime}(\theta_1, \ldots, \theta_p)^*]$ and $t_{j, I}^{\alpha \alpha^\prime}(\theta_1, \ldots, \theta_p) \equiv {\rm Im}[t_j^\alpha(\theta_1, \ldots, \theta_p)t_j^{\alpha^\prime}(\theta_1, \ldots, \theta_p)^*]$. 
Eq.~\eqref{quadratic_E} can be rewritten as:
\begin{align}
\label{E_functional}
    E(\theta_1, \ldots, \theta_p) = \sum_{\alpha \leq \alpha^\prime} \left\{t_{j, R}^{\alpha \alpha^\prime}(\theta_1, \ldots, \theta_p) {\rm Re}[\tilde{H}^{\alpha \alpha^\prime}] + t_{j, I}^{\alpha \alpha^\prime}(\theta_1, \ldots, \theta_p) {\rm Im}[\tilde{H}^{\alpha \alpha^\prime}] \right\} \text{,}
\end{align}
where $t_{j, I}^{\alpha \alpha} = 0$ and $\tilde{H}^{\alpha \alpha} = 0$.
Every set of parameters $(\theta_1, \ldots, \theta_p)$ defines a circuit whose energy can be measured with circuits of depth $d$, $E(\theta_1, \ldots, \theta_p)$, and lets us write a linear equation in the unknowns ${\rm Re}[\tilde{H}^{\alpha \alpha^\prime}]$ and ${\rm Im}[\tilde{H}^{\alpha \alpha^\prime}]$.
Given $D$ pairs $((\theta_1, \ldots, \theta_p), E(\theta_1, \ldots, \theta_p))$, one generates a set of $D$ linear equations.
The maximum number of linearly independent equations one can generate is equal to the number of linearly independent functions in the set $T_{j} \equiv \left\{ t^{\alpha \alpha^\prime}_{j, R}(\theta_1, \ldots, \theta_p) \right\} \cup \left\{ t^{\alpha \alpha^\prime}_{j, I}(\theta_1, \ldots, \theta_p) \right\}$; we denote this as the \emph{rank} of $T_j$.
This rank can be easily computed by evaluating the functions in $T_j$ over generic and randomly chosen points $(\theta_1, \ldots, \theta_p)$ and computing the rank of the set of vectors whose entries are the evaluated functions.

In principle, one should be able to reconstruct $\tilde{H}$ by solving a system with enough equations.
In practice, if the rank of $T_j$ is smaller than the number of entries in $\tilde{H}$ (note that this is equal to the number of independent, real parameters in $\tilde{H}$), then we cannot fully reconstruct $\tilde{H}$, since we have a non-unique solution to our system of equations. However, any $\tilde{H}$ compatible with these equations yields the correct energy function $E(\theta_1, \ldots, \theta_p)$.
As an example, a single $U(4)$ two-qubit gate could in principle require of 256 equations to fully characterize the energy function $E$, given that $\tilde{H}$ is a $16 \times 16$ Hermitian matrix.
However, the rank of $T_j$ is 226, and so only 226 equations are needed to obtain $E(\theta_1, \ldots, \theta_p)$.
As a second example, in the case of a rotation one-qubit gate over a fixed plane in the Bloch sphere, one only needs 3 equations in order to determine $E$, and our method reproduces the 3-parameter quadrature method of Refs.~\cite{nakanishi_sequential_2020,parrish_jacobi_2019}.

Considerations about the rank of $T_j$ are important in order to determine the $\emph{minimum}$ number of equations needed. In practice, by increasing the number of equations to determine $E(\theta_1, \ldots, \theta_p)$ and partially determine $\tilde{H}$ by linear least squares over an over-constrained system of equations, we can add additional robustness into our calculation.

\end{document}